\documentclass[11pt]{article} 
\usepackage{jheppub}

\usepackage{amsmath,amsfonts,amsbsy,amssymb,array,accents,dsfont}
\usepackage{enumerate}
\usepackage[shortlabels]{enumitem}
\usepackage{graphicx} 
\usepackage{subcaption} 
\usepackage{upgreek}
\usepackage{bm}
\usepackage{slashed}

\let\includefigures=\iftrue
\let\useblackboard==\iftrue
\newfam\black

\usepackage[dvipsames]{xcolor}

\definecolor{myblue}{RGB}{85,130,255}%{55, 100, 210}
\definecolor{myred}{RGB}{200, 45, 40}

\PassOptionsToPackage{ 
     colorlinks=true,
     linkcolor=myBlue,
     urlcolor=blue,
     filecolor=black,
     citecolor=myRed,
}{hyperref}

\usepackage{xparse}
\usepackage{xspace}

\NewDocumentCommand\eqn{om}{%
  \IfNoValueTF{#1}
     {\[ #2 \]}
     {\begin{equation}\label{#1} #2  \end{equation} \expandafter\newcommand\csname #1\endcsname{\eqref{#1}\xspace}\ignorespaces}
}
\NewDocumentCommand\eqna{om}{%
  \IfNoValueTF{#1}
    {\begin{align*} #2 \end{align*}}
    {\begin{equation}\label{#1}\begin{split} #2  \end{split}\end{equation} \expandafter\def\csname #1\endcsname{\eqref{#1}\xspace}\ignorespaces}
}

\newcommand{\rcite}{\cite}

\def\xx{{\bf x}}

\def\prop{{\mathfrak {log}}}

\def\sl{\text{sl}}

\def\sltwo{\ensuremath{SL(2,\bR)}}

\def\sutwo{{SU(2)}}

\def\tight#1{\! #1 \!}  % tightens annoying spacing in equations

\def\({\left(}
\def\){\right)}
\def\[{\left[}
\def\]{\right]}

\def\eg{{e.g.}}
\def\cf{{c.f.}}

\def\gstrsq	{g_{\textit s}^{2}}

\def\nfive{{n_5}}

\def\flabel{{\sst (m)}}

\def\A{{\mathsf A}}
\def\B{{\mathsf B}}
\def\C{{\mathsf C}}

\def\sfA{{\mathsf A}}

\DeclareMathSymbol{\medhatsym}{\mathord}{largesymbols}{"62} % basic symbol

\DeclareMathSymbol{\medtildesym}{\mathord}{largesymbols}{"65}% basic symbol

\makeatletter
\newcommand*\rel@kern[1]{\kern#1\dimexpr\macc@kerna}
\newcommand*\widebar[1]{%
  \begingroup
  \def\mathaccent##1##2{%
    \rel@kern{0.8}%
    \overline{\rel@kern{-0.8}\macc@nucleus\rel@kern{0.2}}%
    \rel@kern{-0.2}%
  }%
  \macc@depth\@ne
  \let\math@bgroup\@empty \let\math@egroup\macc@set@skewchar
  \mathsurround\z@ \frozen@everymath{\mathgroup\macc@group\relax}%
  \macc@set@skewchar\relax
  \let\mathaccentV\macc@nested@a
  \macc@nested@a\relax111{#1}%
  \endgroup
}
\makeatother

\def\One{{\hbox{1\kern-1mm l}}}

\def\barray{\begin{array}}
\def\earray{\end{array}}
\def\be{\begin{equation}}
\def\ee{\end{equation}}
\def\bea{\begin{align}}
\def\eea{\end{align}}
\def\bal{\begin{align}}
\def\eal{\end{align}}
\def\nn{\nonumber}

\newcommand{\bR}{{\mathbb R}}
\newcommand{\bS}{{\mathbb S}}
\newcommand{\bT}{{\mathbb T}}

\newcommand{\bZ}{{\mathbb Z}}

\definecolor{cardinal}{rgb}{0.6,0,0}
\definecolor{darkgreen}{rgb}{0,0.4,0}
\definecolor{green}{rgb}{0,0.4,0}
\definecolor{golden}{rgb}{0.92, 0.7, 0}
\definecolor{midnight}{rgb}{0, 0, 0.5}
\definecolor{darkblue}{rgb}{0, 0, 0.7}

\numberwithin{equation}{section}

\mathchardef\mhyphen="2D

\def\cA{\mathcal {A}}

  \def\cO{\mathcal {O}}

\def\cV{\mathcal {V}}

\def\one{{\hbox{\kern+.5mm 1\kern-.8mm l}}}
\def\zero{{\hbox{0\kern-1.5mm 0}}}

\def\id{\textrm{id}}

\def\id{{1 \kern-.28em {\rm l}}}

\def\journal#1&#2(#3){\unskip, \sl #1\ \bf #2 \rm(19#3) }
\def\andjournal#1&#2(#3){\sl #1~\bf #2 \rm (19#3) }

\def\eg{{\it e.g.}}
\def\cf{{\it c.f.}}

\def\sst{\scriptscriptstyle}

\def\vev#1{\langle#1\rangle}

\def\One{{1\hskip -3pt {\rm l}}}

\catcode`\@=11
\def\slash#1{\mathord{\mathpalette\c@ncel{#1}}}
\def\underrel#1\over#2{\mathrel{\mathop{\kern\z@#1}\limits_{#2}}}

\catcode`\@=12

\def\vev#1{\left\langle #1 \right\rangle}

\def\eg{{\it e.g.}}

\title{%\LARGE 
{
BPS Fivebrane Stars II:
Fluctuations
}}

\author{Emil J. Martinec$^a$ {\it and} Yoav Zigdon$^{b}$\\}

\affiliation[a]{
Kadanoff Center for Theoretical Physics, Enrico Fermi Institute, and Department of Physics\\ 
University of Chicago\\ 
5640 S. Ellis Ave.\\
Chicago IL 60637, USA\\ 
}

\affiliation[b]{
Department of Applied Mathematics and Theoretical Physics\\
University of Cambridge\\
Cambridge, CB3 0WA, United Kingdom \\
}

 \emailAdd{e-martinec@uchicago.edu}
 \emailAdd{ yz910@cam.ac.uk}

\abstract{
We investigate quantum fluctuations of metric components in coherent 1/2-BPS bound states of $n_1$ fundamental strings and $n_5$ NS5-branes.  The leading order contribution in an expansion in $1/(n_1n_5)$ is calculated via a combination of analytical and numerical methods.
We find that the fluctuations are small away from a tiny distance from the source, comparable to the 6d Planck scale.  Comparing this result with an analysis in the literature of fluctuations in the maximally mixed state, we conclude that the large fluctuations previously found for the latter are statistical rather than quantum in nature, and that perturbative string theory provides an accurate description of these backgrounds. 
}

\begin{document}
\hypersetup{pageanchor=false}
%\begin{titlepage}
\maketitle
%\thispagestyle{empty}
%\end{titlepage}
\hypersetup{pageanchor=true}
\pagenumbering{arabic}

%\toc
%\thispagestyle{empty}

%\vskip 1cm
%\hrule

%%%%%%%%%%%%%%%%%%%%%%%%%%%%%%%%%%%%%%%%%%%%%%%%%%%%%%%%%%%%%%%%
%%%%%%%%%%%%%%%%%%%%%%%%%%%%%%%%%%%%%%%%%%%%%%%%%%%%%%%%%%%%%%%%

%%%%%%%%%%%%%%%%%%%%%%%%%%%%%%%%%%%%%%%%%%%%%%%%%%%%%%%%%%%%%%%%
%%%%%%%%%%%%%%%%%%%%%%%%%%%%%%%%%%%%%%%%%%%%%%%%%%%%%%%%%%%%%%%%

%\begin{enumerate}[start=1,
%    labelindent=\parindent,
%    leftmargin =2.5\parindent,
%    label=(WII-\arabic*)]
%\item
%\label{WII-1}
%stuff
%\item
%\label{WII-2}
%more stuff
%\end{enumerate}

%Later on I want to refer to \ref{WII-1}

%%%%%%%%%%%%%%%%%%%%%%%%%%%%%%%%%%%%%%%%%%%%%%%%%%%%%%%%%%%%%%%%
%%%%%%%%%%%%%%%%%%%%%%%%%%%%%%%%%%%%%%%%%%%%%%%%%%%%%%%%%%%%%%%%

\section{Introduction and summary} 
\label{sec:intro}
What is the strength of quantum fluctuations in the brane bound states encountered in holography?  
Small relative quantum fluctuations of observables, together with weak string coupling and slowly-varying fields in string units, allow one to describe the structure of the brane bound states using supergravity. In contrast, order one or large relative quantum fluctuations of observables imply that the semi-classical approximation breaks down and another description must be used instead.

A relatively simple subset of brane bound states are those that preserve 8 supercharges of type II superstring theory compactified on $\bT^5$ and carry onebrane and fivebrane charges. Taking a near-source limit results in a geometry $AdS_3\times\bS^3\times \bT^4$; string theory in this limit is dual to a 2d CFT.  BPS spectra and correlation functions on the two sides have been shown to match, providing ample evidence for the correspondence. 

In this paper we focus on 1/2-BPS bound states of $n_5$ Neveu-Schwarz fivebranes (NS5) that wrap $\bT^4 \times \bS^1 _y$ and $n_1$ fundamental strings (F1) that wrap the $\bS^1 _y$ factor. The Lunin-Mathur supergravity solutions~\cite{Lunin:2001fv,Kanitscheider:2007wq} describe the geometry sourced by such bound states. Each bound state is specified by a set of profile functions $f^I(v)$ where $v=t+y$ ($t$ is Lorentzian time and $y$ parametrizes the spatial circle $\bS^1 _y$).  As we review in section 2, the metric, dilaton and potentials sourced by the bound state are given in terms of harmonic functions and forms determined in terms of $f^I(v)$.

We will also be interested in the T-dual NS5-P frame, where momentum waves propagate on the fivebranes along $\tilde\bS^1_y$; $f^I(v)$ specifies the wave profile, four of whose polarizations specify the location of the fivebrane in its transverse space.  It was shown in~\rcite{PaperA} that this duality frame is more appropriate for the description of the background near the brane source.

The phase space of classical solutions~\cite{Rychkov:2005ji} admits a geometric quantization in which the Fourier mode amplitudes of the $f^I(v)$ become creation/annihilation operators in the mode Fock space; the classical solutions are coherent states in this Hilbert space. The entropy of 1/2-BPS configurations was shown to scale like the logarithm of the dimension of a superselection sector of the Hilbert space~\cite{Rychkov:2005ji,Krishnan:2015vha} (see also~\cite{CabreraPalmer:2004asc}).

The Hilbert space structure allows one to superpose states, corresponding to the ability to superpose the harmonic functions that specify the geometry. 
In the limit of small chemical potential, a solution corresponding to the averaged harmonic functions was found in~\cite{Alday:2006nd,Balasubramanian:2008da,Raju:2018xue}, reviewed and further explored in a companion paper~\rcite{PaperA} to the present work. The expectation values of these and other observables in typical states are well-approximated by the expectation values in the ensemble. 

The NS5-F1 ``ensemble geometry'' looks like the extremal BTZ black hole until one gets to a radial position where one must switch to the NS5-P frame to have a valid effective description.  The spatial $\bS^1_y$ circle of the NS5-F1 background shrinks with decreasing radius until it reaches the string scale, and in the T-dual NS5-P frame the $\tilde\bS^1_y$ circle grows with decreasing radius; in this latter frame, the circle size saturates at a large value in the core of the source.  The geometry at the source has the structure of a spherically symmetric blob in the four transverse dimensions.  

The size of the angular $\bS^3$ transverse to the fivebranes is fixed at the AdS length scale $\sqrt{\nfive\alpha'}$ outside the blob, supported at that size by the magnetic NS-NS flux threading it; inside the blob, the flux gradually turns off as more and more of the fivebrane source lies outside a given radius, and the angular sphere decreases smoothly to zero size at the origin where the flux turns off.  

Time redshifts as one approaches the matter sources, but saturates at a large finite value. The blob should not be thought of as a black hole, because it is smooth and differs from the extremal black hole geometry over a region much bigger than the scale of the ``stretched horizon'' where the Bekenstein-Hawking entropy is comparable to the entropy of the BPS ensemble.  Instead, the microstates are those of a ``fivebrane star''.  

Another application of the Hilbert space structure of the 1/2-BPS states is the calculation of quantum fluctuations in observables built out of the profiles $f^I(v)$.  The consideration of fluctuations restricted to the 1/2-BPS configuration space is strictly speaking only valid for BPS observables for which fluctuations outside this subspace cancel due to supersymmetry.  The two-point correlators of harmonic functions in 1/2-BPS states we consider here in principle do receive contributions from non-BPS intermediate states; we will nevertheless proceed under the assumption that these contributions are sub-dominant to the contributions of 1/2-BPS intermediate states which we evaluate below.

In the context of the D1-D5 system in a state which can be interpreted as the grand-canonical ensemble of 1/2-BPS states, the authors of~\cite{Raju:2018xue} calculated the relative fluctuations of the harmonic functions, and found that they are order one even ``well outside'' where the fuzzy bound state is quasi-localized.  By ``well outside'' we mean at an invariant distance of a few times the blob size ($\sqrt{n_5 \alpha'}$) away from the center of the bound state.  The order one relative fluctuations together with their result that the size of the $\bS^1 _y$ falls below the 10D Planck scale, led them to conclude that supergravity is invalid for describing typical states, and that the one solution which reliably describes an ensemble of typical states is the zero-mass and non-rotating BTZ black hole multiplied by the angular three-sphere and the compactification four-torus. 

However, it is not clear directly from the analysis of~\cite{Raju:2018xue} whether their result follows from the bound state being intrinsically quantum, or whether it is rather due to their special choice of a state which is approximately the maximally mixed state of given charges.%
\footnote{The ensemble of fixed charge is the maximally mixed state;~\cite{Raju:2018xue} worked in the ensemble of fixed chemical potential.}

In other words, are generic NS5-F1 bound states highly quantum by their very nature, or are the fluctuations observed in~\cite{Raju:2018xue} better interpreted as statistical fluctuations resulting from the use of the maximally mixed state?
For example, if we consider the mixed state of a macroscopic object summed over microstates having a variety of positions and orientations (and perhaps different states of its internal degrees of freedom), it will exhibit large fluctuations of various observables, but these are classical statistical fluctuations of the mixed ensemble rather than intrinsically quantum fluctuations of the object.
Our results below support the latter proposition.

\begin{figure}
\centering
\includegraphics[scale=1]{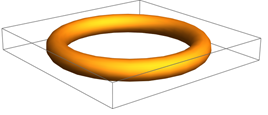}
\caption{An illustration of ring-shaped source composed of rotating NS5-F1 matter. We ask what are the quantum fluctuations of the metric produced by this bound state normal to its worldvolume.}
\label{fig:ring}
\end{figure}

We will consider quantum fluctuations of the metric components of ring-shaped $(AdS_3\times\bS^3)/\bZ_n$ Lunin-Mathur solutions, of the sort depicted in Figure~\ref{fig:ring}. These solutions correspond via the map of~\rcite{Rychkov:2005ji,Kanitscheider:2007wq} to coherent states in the restricted phase space of 1/2-BPS solutions.  
We find that the relative quantum fluctuations in these states are suppressed by an inverse power of the central charge $c=6n_1 n_5$ away from the source, and become of order one at a distance scale comparable to but somewhat larger than the six-dimensional Planck length from the ring. They are also enhanced by~$n$, the orbifold parameter that labels the coherent state, making fluctuations large when the orbifold parameter is of order the central charge. We conclude that there is an overcomplete basis of coherent states for which the semi-classical description is applicable beyond of order Planck-scale distance from the brane sources, in any individual coherent basis state.  Furthermore, in a typical state the source is spread out, with fivebrane strands well-separated relative to the 6d Planck scale~\rcite{Mathur:2007sc,PaperA}; thus any individual element of the maximally mixed ensemble has small fluctuations except very near the source, and so the fluctuations observed in~\rcite{Raju:2018xue} are statistical fluctuations of the source rather than quantum fluctuations of the geometry.

The paper is organized as follows. In section~\ref{sec:flucts} we review the map between bulk microstates, geometries and CFT states. This map is used to review a calculation of the expectation values of harmonic functions in the coherent states that give rise to the circular Lunin-Mathur solution \cite{Lunin:2001fv}. We then discuss the length scales appearing in the problem. This is followed in section~\ref{sec:correlations} by a semi-analytical calculation of the quantum fluctuations of metric components in transverse space.  A discussion completes the paper in section~\ref{sec:disc}.

%%%%%%%%%%%%%%%%%%%%%%%%%%%%%%%%%%%%%%%%%%%%%%%%%%%%%%%%%%%%%%%%
%%%%%%%%%%%%%%%%%%%%%%%%%%%%%%%%%%%%%%%%%%%%%%%%%%%%%%%%%%%%%%%%

\section{Review of one-point functions}
\label{sec:flucts}

\subsection{States in the dual CFT} 
\label{sec:holomap}

The BPS states of the NS5-F1 system are simple to describe in terms of the T-dual along NS5-P system.  There, they are simply BPS waves on the fivebranes.  In a sector where the fivebranes are twisted into a single fivebrane wrapping $n_5$ times around $\bS^1_y$, momenta are fractionated by a factor of $n_5$; one can then have any number $n_k^I$ of modes with momenta $k/N$ in any of 8 bosonic and 8 fermionic polarizations $I$, subject to the overall constraint 
\be
\label{sumrule}
\sum_{k,I} k \, n_k^I = n_1n_5\equiv N  ~.
\ee
We can thus label the 1/2-BPS spectrum via
\be
\label{halfbpsstate}
\big| \{n_k^I\}\big\rangle ~,
\ee
and this labeling passes through the T-duality to the NS5-F1 frame where the labels refer to a collection of (generically fractional, unless $k$ is a multiple of $n_5$) winding strings carried as fivebrane excitations.  The bosonic excitations split into the four scalars $X^{\alpha\dot\alpha}$ that describe the gyration of the fivebrane in its transverse space, and four more comprising a gauge multiplet on the fivebrane.%
\footnote{For type IIB in the NS5-F1 frame, the polarization labels refer to the gauge multiplet $A^{AB}$ on the T-dual fivebrane consisting of a scalar and a self-dual antisymmetric tensor.  The scalar $A^{[AB]}$ is typically referred to in the literature as the ``00'' mode.}
The charges carried by each microstate are given by
\begin{equation}
	Q_1=\frac{g_s^2 n_1 (\alpha')^3}{V_4} ~~,~~~~ Q_5=n_5\alpha' 
\end{equation}
where $g_s$ is the asymptotic string coupling, 
$(2\pi)^4 V_4$ is the volume of the $\bT^4$, 
$n_1$ the number of fundamental strings and $n_5$ the number of NS fivebranes.
The $\bZ_ n$ orbifold geometries are dual to the states
\be
\label{Zkstates}
\Big| \big\{n_ n^{++}\tight=N/ n, {\it others} \tight=0 \big\} \Big\rangle ~.
\ee

The 1/2-BPS supergravity solutions of the NS5-F1 system can be put in a standard form~\cite{Lunin:2001fv,Kanitscheider:2007wq} (restricting for simplicity to solutions with pure NS-NS fluxes; for the general solution, see \eg\ Appendix B of~\cite{Martinec:2022okx})
\begin{align}
\begin{aligned}
\label{LMgeom}
ds^2 &\;=\; \frac{1}{H_1}\bigl[ -(dt+\A)^2 + (dy + \B)^2 \bigr] + H_5 \, d\xx\!\cdot\! d\xx + d|\vec{z}|^2
\\[.2cm]
B &\;=\; 
\frac{1}{H_1} \bigl(dt + \A\bigr)\wedge\bigl(dy+\B\bigr)  + \C_{ij}\, dx^i\wedge dx^j 
\\[.2cm]
e^{2\Phi} &\;=\; 
{\gstrsq}\,\frac{H_5}{H_1}  \,, \qquad~~~ d\C =  *_{\sst\perp} dH_5  
\qquad~~ d\B =- *_{\sst\perp} d\A 
\end{aligned}
\end{align}
where $\xx$ are Cartesian coordinates on the transverse space to the fivebranes, related to $\sltwo\times\sutwo$ Euler angles via
\be
\label{bipolar}
x^1+ix^2 \equiv x^{++} = \cosh\rho\,\sin\theta \, e^{i\phi}
~~,~~~~
x^3+ix^4 \equiv x^{-+} = \sinh\rho\,\cos\theta\, e^{i\psi}  ~.
\ee
The $\vec{z}$ coordinates parametrize the four-torus.

The harmonic forms and functions appearing in this solution can be written in terms of a Green's function representation,
which in the $AdS_3$ decoupling limit takes the form%
\footnote{Unless stated otherwise, here and below the notation $\left|~\right|^2$ means the norm squared of a vector and not a quantity times its complex conjugate. }
\begin{align}
\begin{aligned}
\label{greensfn}
H_5 (\xx) \,=\, \frac{Q_5}{L}\int\limits_0^{L} \frac{dv}{|\xx-\mathbf{f}(v)|^2} ~~, & \qquad~
H_1 (\xx)   \,=\,  
\frac{Q_5}{L}\int\limits_0^{L} \frac{d v ~ \dot{ \mathbf{f}} \tight\cdot \dot{\mathbf{f}}}{|\xx-\mathbf{f}(v)|^2} \;,\\[.2cm]
\A \,=\, \A_{\alpha\dot\alpha} dx^{\alpha\dot\alpha} ~~, \qquad
\A_{\alpha\dot\alpha} \,=\, & -\frac{Q_5}{L} \int\limits_0^{L} \frac{d v \,\dot {\mathbf{f}}^{\alpha\dot\alpha}(v)}{|\xx-\mathbf{f}(v)|^2}
\end{aligned}
\end{align}
involving source profile functions $\mathbf{f}^{\alpha\dot\alpha}(v)$ that describe the locations of the fivebranes in their transverse space, using bispinor indices (overdots denote derivatives with respect to $v$). The parameter $L$ is defined in terms of the number of fivebranes and the asymptotic radius of the y-circle
\begin{equation}
L = \frac{2\pi Q_5}{R_y}.
\end{equation}
The single profile function $\mathbf{f}^{\alpha \dot\alpha} (v)$, can be viewed as resulting from bundling $n_5$ distinct fivebrane profile functions together.
Labeling these individual profiles by $m$, one can then choose twisted boundary conditions for the source profile functions,
\be
\label{twistedbc}
\mathbf{f}^{\alpha\dot\alpha}_\flabel(\tilde{v}+2\pi)=\mathbf{f}^{\alpha\dot\alpha}_{\sst (m+1)}(\tilde{v})  ~,~ \tilde{v}= \frac{2\pi v}{L}
\ee
that bind all the fivebranes together (provided $m$ and $n_5$ are relatively prime) and introduce the fractional moding described above. 
The key point here is that the Fourier amplitudes $\alpha_k^I$ of the source profile functions 
\be
f^I(\tilde v) = \frac{i\mu}{\sqrt{2}}\sum_{k\neq0} \frac{\alpha_k^I}{k}\, e^{ik \tilde{v}}
\ee
are coherent state parameters whose absolute squares equal to $k n_k^I$, with $n_k ^I$ denoting the occupation number of the state with polarization label $I$ and winding number $k$.  We omit a zero mode contribution to the transverse profile functions; the zero mode is non-dynamical in the decoupling limit.

The parameter $\mu$, given by
\begin{equation}
 \mu= \frac{g_s (\alpha')^2}{R_y  \sqrt{V_4}} ~,
\end{equation}
is the effective string scale of an NS5-brane wrapped on $\bT^4$, after a T-duality on $\bS^1_y$.  The functions $f^I(v)$ are waves on that effective string in the T-dual NS5-P frame.

\begin{figure}[ht]
\centering
\includegraphics[width=0.4\textwidth]{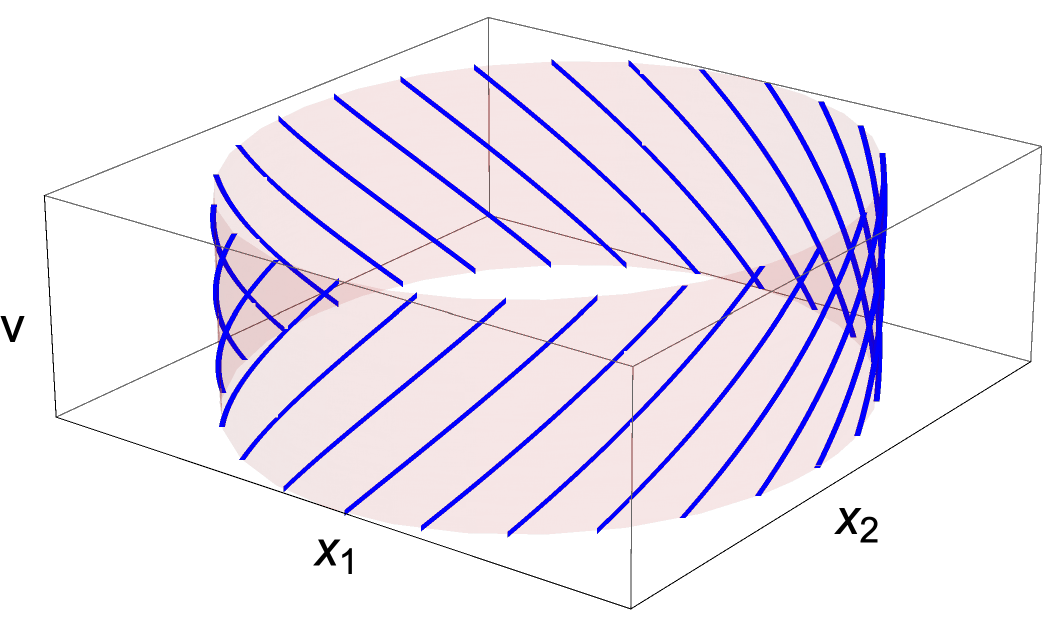}
\caption{\it Circular supertube source profile, in which only a single mode is excited (in this case, $ n = 3$ and $n_5 = 25$), so that the fivebranes spiral around a torus in $(y, x^1, x^2)$ shaded in pink. }
\label{fig:STspiral}
\end{figure}

We thus have a direct map between the labels of 1/2-BPS states and the bulk geometries they correspond to, at the fully non-linear level.  In particular, for the orbifold geometries $(AdS_3\times\bS^3)/\bZ_n$, we know exactly what the fivebranes are doing~-- the source profile function has only a single mode excited
\be
\label{Zkprofile}
f^{++}(\tilde v) = \frac{i\mu}{\sqrt{2}}\frac{\alpha^{++}_ n}{ n} \, e^{i n  \tilde{v}} 
\ee
with 
\be
\label{totallevel}
|\alpha^{++}_ n| = \sqrt{N}  ~,
\ee
and describes fivebranes sitting at $\rho=0$, spiraling around in the torus parametrized by the T-dual to the $AdS_3$ azimuthal coordinate $\sigma=\frac{y}{R_y}$ and the $\bS^3$ Euler angle $\phi$ (see for instance~\cite{Martinec:2017ztd}).  The spiral runs along the $( n,n_5)$ cycle of this torus; see figure~\ref{fig:STspiral}.

As we review in the next subsection, the harmonic functions resulting from \eqref{Zkprofile} are given by (converting from bispinor to vector indices for the one-form $\sfA$)

\begin{align}
\label{circharmfns}
	H_1 (\vec{x}) &= \frac{Q_1}{\sqrt{(|\vec{x}|^2 +a^2)^2-4a^2(x_1 ^2 + x_2 ^2)}}~,
\nn\\[.2cm]
A_1 &= \cA \,x_2 ~~,~~~~ A_2=-\cA \,x_1 ~~,
\nn\\[.1cm]
\cA &= \frac{2\sqrt{Q_1 Q_5} }{\sqrt{(|\vec{x}|^2 +a^2)^2-4a^2(x_1 ^2 + x_2 ^2)} \left(|\vec{x}|^2 +a^2 + \sqrt{(|\vec{x}|^2 +a^2)^2-4a^2(x_1 ^2 + x_2 ^2)}\right)}~,
\nn\\[.2cm]
A_3 &= A_4=0 ~~,
\nn\\[.3cm]
	H_5 (\vec{x}) &= \frac{Q_5}{\sqrt{(|\vec{x}|^2 +a^2)^2-4a^2(x_1 ^2 + x_2 ^2)}}~.
\end{align}

%
%%%%%%%%%%%%%%%%%%

The CFT dual is often described in the language of the symmetric product orbifold, which pertains to a weak-coupling region of the moduli space.  In the symmetric product, 1/2-BPS states are associated to conjugacy classes of the symmetric group, which are labeled by the same data~\eqref{sumrule}, \eqref{halfbpsstate}, and describe collections of copies (cycles) of the block $\bT^4$ CFT that are sewn together by a cyclically twisted boundary condition analogous to~\eqref{twistedbc}.  Each cycle has a collection of 1/2-BPS ground states labeled by the same data as the polarization labels $I$ carried by the bulk fivebrane excitations.

The BPS states are preserved under the marginal deformation to the strongly-coupled regime of the CFT where the bulk dual has a supergravity approximation.  The analysis of~\cite{Martinec:2020gkv,Martinec:2022okx} shows that much of the symmetric product structure survives this deformation.  In particular, the effect of 1/2-BPS string vertex operators is to deform perturbatively the string winding condensate carried by the fivebranes, and at the same time the geometry that the condensate is sourcing.%  
\footnote{The aspects of the vertex operator that are responsible for these two effects are related by FZZ duality~\cite{Giveon:2016dxe,Martinec:2020gkv}.}
For instance, a 1/2-BPS graviton vertex operator $\cV^{\alpha\dot\alpha}_{j',w_y}$ sews together a number of background strings into a longer string, while changing its polarization state:
\be
\label{Vtransition}
\big(|\tight++\rangle_ n\big)^{2j'+1}~\longrightarrow~ |\alpha\dot\alpha\rangle_{(2j'+1) n+w_y n_5}
\ee
where $|I\rangle_k$ denotes a cycle of length $k$ in polarization state $I$
(see~\cite{Martinec:2020gkv,Martinec:2022okx} for details).
Exponentiating the vertex operators thus coherently changes the winding condensate carried by the fivebranes as specified by the profile functions $f^I(\tilde v)$.

%%%%%%%%%%%%%%%%%%%%%%%%%%%%%%%%%%%%%%%%%%%%%%%%%%%%%%%%%%%%%%%%
%%%%%%%%%%%%%%%%%%%%%%%%%%%%%%%%%%%%%%%%%%%%%%%%%%%%%%%%%%%%%%%%
\subsection{Harmonic functions for the single-mode source}
\label{subsec:HarmFnsReview}

The aim of this subsection is to calculate the one-point function of components of the metric induced by circular bound states where a single mode is excited. We choose to introduce a cutoff on the possible winding number, denoted by $\Lambda$, because the integrals we encounter are convergent in this case, and later take a large-cutoff limit.   
We start by writing down the mode expansion for the profile vector:
\begin{equation}\label{ModeExpansion}
	\vec{f} (z) =  \frac{i \mu}{\sqrt{2}}\sum_{m=1} ^{\Lambda}\left( \frac{1}{m} \frac{1}{z^m}\vec{\hat{\alpha}}_m -\frac{1}{m} \vec{\hat{\alpha}}_{-m}  z^m \right) ~~, ~~~~z=e^{i\frac{2\pi v}{L}}.
\end{equation}
We will later take the limit $\Lambda, N \to \infty$ while holding the radius of the ring fixed.
The mode operators satisfy the commutation relations
\begin{equation}
\big[\alpha^j_{n}, \alpha^i_{-m}\big] = n \delta_{n,m}\delta^{i,j}
\end{equation}
where the indices $i,j$ run over the dimensions of transverse space.

We describe the source profile~\eqref{Zkprofile} in terms of coherent states for each mode, with all but the $n^{\rm th}$ mode in their ground states, and the $n^{\rm th}$ modes having coherent state parameters $\vec\lambda_n$ such that the expectation value of the occupation number is $\sqrt{N}$, see eq.~\eqref{totallevel}.  The harmonic functions appearing in the metric are then given by expectation values such as
\begin{equation}\label{StartingPoint}
\big\langle \vec{\lambda}_n \big| \hat{H}_5 (\vec{x}) \big|\vec{\lambda}_n \big\rangle =Q_5 \Big\langle \vec{\lambda}_n \Big| \oint \frac{dz}{2\pi i z}  \frac{1}{|\vec{x}-\vec{f}(z)|^2} \Big|\vec{\lambda}_n \Big\rangle~.
\end{equation}
The following identity is useful:
\begin{equation}\label{Fourierofx-2}
	\frac{1}{|\vec{x}|^2} = \frac{1}{4\pi^2}\int d^4 k_1 \frac{e^{i \vec{k}_1\cdot \vec{x}}}{|\vec{k}_1|^2}~.
\end{equation}
Plugging eq.~(\ref{Fourierofx-2}) into eq.~(\ref{StartingPoint}) results in
\begin{equation}\label{H5Expression}
\big\langle \vec{\lambda}_n \big| \hat{H}_5 (\vec{x})\big|\vec{\lambda}_n \big\rangle =Q_5\oint \frac{dz}{2\pi i z}\frac{1}{(2\pi)^2}\int \frac{d^4 k_1}{|\vec{k}_1|^2} e^{-i \vec{k}_1 \cdot \vec{x}} \big\langle \vec{\lambda}_n \big|  e^{i\vec{k}_1 \cdot \vec{f}}  \big|\vec{\lambda}_n \big\rangle~.
\end{equation}
The exponential $e^{i\vec{k}_1 \cdot \vec{f} (z_1)}$ is regulated by truncating the mode expansion~(\ref{ModeExpansion}) and applying the Baker-Campbell-Hausdorff formula:
\begin{equation}
	 e^{i\vec{k}_1 \cdot \vec{f}(z_1)}\! \; = \prod_{m=1} ^{\Lambda} e^{-\frac{\mu^2 |\vec{k}_1|^2}{4m}}e^{ \frac{\mu}{\sqrt{2m}}z_1 ^m \vec{k}_1 \cdot \hat{a}_m ^{\dagger} }e^{- \frac{\mu}{\sqrt{2m}}\frac{1}{z_1 ^m} k_1 \cdot \hat{a}_m}~.
\end{equation}
In the large $\Lambda$ approximation, one has
\begin{equation}
\log(\Lambda) = \sum_{m=1} ^{\Lambda} \frac{1}{m}.
\end{equation}
The coherent state is an eigenstate of the annihilation operator, thus
\begin{align}\label{Exponential}
	\Big\langle \vec{\lambda}_n \Big|e^{i\vec{k}_1 \cdot \vec{f}(z_1)} \Big|\vec{\lambda}_n\Big\rangle = e^{-\frac{\mu^2 |\vec{k}_1|^2}{4}\log(\Lambda)}e^{ \frac{\mu}{\sqrt{2n}}z_1 ^n \vec{k}_{1} \cdot  \vec{\lambda}_{n} ^*- \frac{\mu}{\sqrt{2n}}\frac{1}{z_1 ^n} \vec{k}_{1}\cdot \vec{\lambda}_{n}}~.
\end{align}
Substituting eq.~(\ref{Exponential}) into eq.~(\ref{H5Expression}) implies that
\begin{align}
\big\langle \vec{\lambda}_n \big| \hat{H}_5 (\vec{x})\big|\vec{\lambda}_n \big\rangle =Q_5\oint \frac{dz}{2\pi i z}\frac{1}{(2\pi)^2}\int \frac{d^4 k_1}{|\vec{k}_1|^2} e^{-i \vec{k}_1 \cdot \vec{x}} e^{-\frac{\mu^2 |\vec{k}_1|^2}{4}\log(\Lambda)}e^{ \frac{\mu}{\sqrt{2n}}z_1 ^n \vec{k}_{1} \cdot  \vec{\lambda}_{n} ^*- \frac{\mu}{\sqrt{2n}}\frac{1}{z_1 ^n} \vec{k}_{1}\cdot \vec{\lambda}_{n}}.
\end{align}
Define
\begin{equation}
	\Delta \vec{x} \equiv \vec{x} - i\frac{\mu}{\sqrt{2n}}\left( \frac{\vec{\lambda}_n}{z^n}-\vec{\lambda}_n ^* z^n\right).
\end{equation}
This is interpreted as the difference between the position vector $\vec{x}$ where the one-point function is evaluated, and the expectation value of the profile vector (\ref{ModeExpansion}) in the coherent state in question. 
Then the following integral emerges
\begin{equation}
	\Big\langle \vec{\lambda}_n \Big| \hat{H}_5 (\vec{x}) \Big|\vec{\lambda}_n \Big\rangle =Q_5\oint \frac{dz}{2\pi i z}~\frac{1}{(2\pi)^2}\int \frac{d^4 k_1}{|\vec{k}_1|^2} e^{-i\vec{k}_1 \cdot \Delta \vec{x}} e^{-\frac{k_1 ^2 \mu^2}{4} \log(\Lambda)}.
\end{equation}	
Performing the $\vec{k}_1$ integrals (e.g. by working in spherical coordinates) leads to
\begin{equation}\label{Result4DGreensFunction}
	\Big\langle \vec{\lambda}_n \Big| \hat{H}_5 (\vec{x}) \Big|\vec{\lambda}_n \Big\rangle =Q_5\oint \frac{dz}{2\pi i z}~\frac{1}{|\Delta \vec{x}|^2}\left(1-e^{-\frac{|\Delta \vec{x}|^2}{\mu^2 \log(\Lambda)}}\right).
\end{equation}	
We now note that as long as $\frac{N}{n^2}\gg 1$, the right term on the R.H.S. of eq.~(\ref{Result4DGreensFunction}) is exponentially small in the large-N limit and hence can be neglected. To this end we employ a useful relation between the parameter $\mu$ and the radius $a$ of the ring associated with the orbifold geometry \cite{Lunin:2001fv}
\begin{equation}
	a= \frac{\sqrt{Q_1 Q_5}}{n R_y} = \frac{g_s \sqrt{n_1 n_5} (\alpha')^2}{nR_y \sqrt{V_4}}=\frac{\mu \sqrt{n_1 n_5}}{n}~. 
\end{equation} 
Note that this definition of $a$ differs by a factor of $n$ from that used in~\rcite{PaperA}.
Further defining
\begin{equation}
	\tilde{x} = \frac{\vec{x}}{a} ~~,~~~~ \tilde{y} = \frac{\vec{y}}{a}~,
\end{equation}
and setting $\Lambda =c N$ with $c$ an order one number, one obtains
\begin{equation}
\frac{|\Delta \vec{x}|^2}{\mu^2 \log(\Lambda)} = \frac{N|\Delta \tilde{x}|^2}{n^2\log(c N)}.
\end{equation}
It follows that in the $\frac{N}{n^2}\gg 1$ regime, the right term in parenthesis on the R.H.S of eq.~(\ref{Result4DGreensFunction}) can be neglected and
\begin{equation}\label{Result4DGreensFn2}
	\Big\langle \vec{\lambda}_n \Big| \hat{H}_5 (\vec{x}) \Big|\vec{\lambda}_n \Big\rangle \approx Q_5\oint \frac{dz}{2\pi i z}~\frac{1}{|\Delta \vec{x}|^2}.
\end{equation}	
One could have normal ordered $H_5 (\vec{x})$, and then the approximation sign in (\ref{Result4DGreensFn2}) would have been replaced by an equality sign.

The norm of the $|\Delta \vec{x}|$ vector squared is given by:
\begin{align}
	&|\Delta \vec{x}|^2 = |\vec{x}|^2 -i\mu \sqrt{\frac{2}{n}}\lambda\left[x\frac{1}{z_1 ^n}-\bar{x}z_1 ^n\right]+\frac{2\mu^2 \lambda^2}{n}
	 = \frac{1}{z_1 ^n}i\mu \sqrt{\frac{2}{n}} \lambda \bar{x} ~ (z_1 ^n-z_+)(z_1 ^n - z_-)~,
\end{align}
where
\begin{equation}
    x=x_1+ix_2 
~~,~~~~ 
    \bar{x}=x_1-ix_2
~~,~~~~
    {w}\equiv |\vec{x}|^2 + \frac{2\mu^2 \lambda^2}{n}
~~,~~~~
	z_{\pm} = \frac{{w} \pm \sqrt{{w} ^2 -8\frac{\mu^2 \lambda^2}{n}x \bar{x} }}{-2i \sqrt{\frac{2\mu^2}{n}} \lambda  \bar{x}} 
 ~~.
\end{equation}
Following~\cite{Bena:2016agb}, the residue theorem yields 
\begin{equation}
\label{zint-ring}
	\oint \frac{d(z_1 ^n)}{2\pi i nz_1 ^n} \frac{1}{|\Delta \vec{x}|^2}=\oint \frac{dz }{2\pi i z}\frac{1}{|\vec{x}|^2 -i \sqrt{\frac{2\mu^2}{n}}\lambda\left[x\frac{1}{z }-\bar{x}z \right]+\frac{2\mu^2 \lambda^2}{n}}=
	\frac{1}{\sqrt{\left(|\vec{x}|^2 + \frac{2\mu^2 \lambda^2}{n}\right)^2 -\frac{8\mu^2 \lambda^2}{n} x\bar{x}}}~.
\end{equation}
Identifying
\begin{equation}\label{a-mu}
	a\equiv \sqrt{\frac{2}{n}}\,\mu\lambda
 = \mu\frac{\sqrt{N}}{n}
 ~,
\end{equation}
one obtains the one-point function~\eqref{circharmfns} of the harmonic function $H_5$ (again recall that the definition $\frac a\mu=\frac{\sqrt{N}}n$ differs from that in~\rcite{PaperA} by the factor $\frac 1n$).
One similarly obtains the expectation values of the other harmonic forms $\sfA$ and $H_1$ given in~\eqref{circharmfns}.

%%%%%%%%%%%%%%%%%%%%%%%
\subsection{Discussion of scales}
\label{subsec:DiscussionOfScales}

We would like to explain the physical meaning of the length scale $\mu$ encountered in the mode expansion of the profile functions. The decoupling limit sends the asymptotic radius of the spatial circle to infinity, implying that $\mu$ tends to zero. However, the proper distance corresponding to the coordinate separation $\mu$ between two points in the vicinity of the radius of the ring remains finite: 
We show that it is comparable to the six-dimensional Planck length scale $\ell_{P,6D}$ times the square root of the orbifold number, $\sqrt{n}$. Choose for example 
\begin{align}
x_2=x_3=x_4=0 ~~&,~~~~ x_1 =a+\mu~,
\nn\\[.2cm]
y_2=y_3=y_4=0 ~~&,~~~~ y_1 =a+2\mu~.
\end{align}
The invariant distance between these two points, denoted by $D$, is given by
\begin{equation}
D=\sqrt{Q_5} \int_{a+\mu} ^{a+2\mu} \frac{dx_1}{\sqrt{x_1 ^2 -a^2}}\approx \sqrt{2n_5 \alpha'} \, \big(\sqrt{2}-1\big) \sqrt{\frac{\mu}{a}}.
\end{equation}
The invariant distance between two points near the ring is approximately independent on whether the points are separated in the radial or angular directions.
Since
\begin{equation}
\frac{\mu}{a}=\frac{n}{\sqrt{n_1 n_5}}~,
\end{equation}
it follows that
\begin{equation}
D\approx \sqrt{n} \sqrt{2}(\sqrt{2}-1)\left(\frac{n_5}{n_1}\right)^{\frac{1}{4}} \sqrt{\alpha'}~. 
\end{equation}
We now calculate the six-dimensional Planck scale. The attraction mechanism fixes the dilaton in the deep interior of the geometry as follows
\begin{equation}\label{asdf}
e^{2\phi} = \frac{n_5 V_4}{n_1 (\alpha')^2}.
\end{equation}
Due to this fixed scalar condition~\eqref{asdf}, the 10-dimensional Planck length is determined from 
%Polchinski Volume 2 page 150 eq. (13.3.22). I also assume the relations 1/(2\kappa^2) = (M_P^{10D}) ^{8}/2  and \ell_P ^{10D} = 1/M_P ^{10D}.
\begin{equation}\label{PlanckScale10D}
2 \left(\ell_{P}^{(10D)}\right) ^8 =(2\pi)^7 \frac{n_5 V_4}{n_1 (\alpha')^2} ~(\alpha')^4~.
\end{equation}
Therefore, the six-dimensional Planck length is given by
\begin{equation}\label{PlanckScale6D}
\left(\ell_P ^{(6D)}\right) ^4 = \frac{(\ell_P ^{(10D)})^8}{(2\pi)^4 V_4} \quad\Longrightarrow\quad  \ell_P ^{(6D)} = \left(4\pi^3\frac{n_5}{n_1}\right) ^{\frac{1}{4}}\sqrt{\alpha'}~.
\end{equation}
Consequently,
\begin{equation}
D\approx  2\pi^{\frac{3}{4}}(\sqrt{2}-1)\sqrt{n}\,\ell_{P}^{(6D)}~.
\end{equation}
%The numerical factor evaluates to 12.2828
For the maximally-spinning supertube, $n=1$ and $D\propto \ell_{P}^{(6D)}$, and perturbative string theory is not valid for describing physics between the positions $\vec{x}$ and $\vec{y}$ defined above.

It is useful to compare this scale with the characteristic length scale of the solution obtained for the grand-canonical ensemble with small chemical potential conjugate to the string charge (the S-dual of the D1-D5 solution written in~\cite{Alday:2006nd,Balasubramanian:2008da,Raju:2018xue}). 
\begin{equation}
\label{ensgeom}
	ds^2 _6 = \frac{r^2}{Q_1\big(1-e^{-{r^2}/{r_b ^2}} \big)}\left( -dt^2 + dy^2\right)+ \frac{Q_5\big(1-e^{-{r^2}/{r_b ^2}} \big)}{r^2}\left(dr^2 + r^2 d\Omega_{3} ^2 \right) ~.
\end{equation}
The following coordinate-distance sets this scale:
\begin{equation}
r_b \propto (n_1 n_5) ^{\frac{1}{4}} \mu~. 
\end{equation}
Given eq. (\ref{PlanckScale6D}), the coordinate distance between the origin $r=0$ and the scale $r_b$ amounts to the invariant distance
\begin{equation}
\label{blobsize}
L \propto  \sqrt{n_5 \alpha'}=\,\ell_{AdS}~.
\end{equation}

%%%%%%%%%%%%%%%%%%%%%%%
\section{Fluctuations in a coherent state}
\label{sec:correlations}

Consider a coherent state $|\vec{\lambda}_{n}\rangle$ that was defined in subsection \ref{subsec:HarmFnsReview}. We would like to calculate 
\begin{align}
	\oint \frac{dz_1}{2\pi i z_1}\oint \frac{dz_2}{2\pi i z_2}\Big\langle \vec{\lambda_{n}}\Big| \frac{1}{|\vec{x}-\vec{f}(z_1)|^2} \; \frac{1}{|\vec{y}-\vec{f}(z_2)|^2} \Big|\vec{\lambda_{n}}\Big\rangle ~.
\end{align}
Identity (\ref{Fourierofx-2}) allows one to write the integrand as
\begin{align}
\label{2ptGreenfn0}
	&\Big\langle \frac{1}{|\vec{x}-\vec{f}(z_1)|^2} \; \frac{1}{|\vec{y}-\vec{f}(z_2)|^2} \Big\rangle =\frac{1}{(2\pi)^4}\int d^4 k_1 \int d^4 k_2 \frac{e^{-i \vec{x}\cdot \vec{k}_1 -i \vec{k}_2\cdot \vec{y}}}{|\vec{k}_1| ^2 |\vec{k}_2| ^2}\Big \langle e^{i \vec{k}_1 \cdot \vec{f}(z_1)} \; e^{i \vec{k}_2 \cdot \vec{f}(z_2)}\Big\rangle~. \nonumber\\
\end{align}

%%%%%%%%%%%%%%%%%%%%%%%
\subsection{Step 1: Two-point function of exponential operators}
\label{subsec:TwoExpVev}

Typically, one would invoke normal ordering of the exponential operators in expression (\ref{2ptGreenfn0}), and proceed to calculate the correlation function as one does in the Koba-Nielsen amplitudes of string theory.  However, we are not here calculating the string S-matrix~-- for instance we do not impose momentum conservation (which would arise if the profile $\vec f$ had a zero mode that we integrated over).  The difference between the exponential operator $exp[i\vec k\tight\cdot{\vec f}\,]$ with and without normal ordering is a Gaussian factor 
\be
e^{-\frac{k^2 \mu^2}{4} \sum_{m=1} ^{\Lambda} \frac{1}{m}} ~.
\ee
where we have cut off the mode sum at mode number $\Lambda$.
The reason that we must work with the exponential of the position operator $\vec f$ rather than some normal ordered version is that these Gaussian factors are essential for rendering the $\vec{k}_1,\vec{k}_2$ integrals convergent (indeed, without them even the two-point correlation functions of the form~\eqref{2ptGreenfn0} for a single harmonic oscillator would diverge).  And in any event, it is functions of the position operator and not some normal ordered version that we are interested in for the problem at hand.

The two-point function of the exponentials is 
\begin{align}
\label{exptl2ptfn}
&\Big\langle \vec{\lambda}_n \Big|  e^{i\vec{k}_1 \cdot \vec{f}(z_1)}\; e^{i\vec{k}_2 \cdot \vec{f} (z_2)} \Big| \vec{\lambda}_n\Big\rangle 
\nn\\
&\hskip 1cm
= \Big\langle \vec{\lambda}_{n} \Big| 
e^{-\frac{\mu^2k_1^2}{4}\log(\Lambda)}\,e^{\frac{\mu}{\sqrt{2n}}z_1 ^n \vec{k}_{1} \cdot \vec{\hat{a}}_{n} ^{\dagger}} e^{-\frac{\mu}{\sqrt{2n}}\frac{1}{z_1 ^n} \vec{k}_{1}  \vec{\hat{a}}_{n}}
\;e^{-\frac{\mu^2 k_2^2}{4}\log(\Lambda)}\,e^{\frac{\mu}{\sqrt{2n}}z_2 ^n \vec{k}_{2} \cdot \vec{\hat{a}}_{n} ^{\dagger} }e^{-\frac{\mu}{\sqrt{2n}}\frac{1}{z_2 ^n} \vec{k}_{2} \cdot \vec{\hat{a}}_{n}} \Big|\vec{\lambda}_{n}\Big\rangle
\nn\\[.2cm]
&\hskip 1cm
=  e^{-\frac{\mu^2 (k_1^2+k_2^2)}{4}\log(\Lambda)} e^{-\frac{\mu}{\sqrt{2n}}\frac{1}{z_2 ^n} \vec{k}_{2} \cdot \vec{\lambda}_{n}} e^{\frac{\mu}{\sqrt{2n}}z_1 ^n \vec{k}_{1}\cdot  \vec{\lambda}_{n} ^*}  \Big\langle \vec{\lambda}_{n}\Big|e^{- \frac{\mu}{\sqrt{2n}}\frac{1}{z_1 ^n} \vec{k}_{1} \cdot \vec{\hat{a}}_{n}}e^{ \frac{\mu}{\sqrt{2n}}z_2 ^n \vec{k}_{2} \cdot \vec{\hat{a}}_{n} ^{\dagger} } \Big|\vec{\lambda}_{n}\Big\rangle~.
\end{align}
One now systematically reorders the exponential operators using
\begin{equation}\label{reorder}
	e^{u \hat{a} } e^{v \hat{a}^{\dagger}} = e^{u\hat{a}+v\hat{a}^{\dagger} +\frac{1}{2}uv }=e^{uv}e^{v \hat{a}^{\dagger}}	e^{u \hat{a} }~,
\end{equation}
with the result
\begin{align}
\label{exptl2ptfn-result} 
& \Big\langle \vec{\lambda}_n \Big| e^{i\vec{k}_1 \cdot \vec{f}(z_1)} \;  e^{i \vec{k}_2 \cdot \vec{f} (z_2)} \Big|\vec{\lambda}_n\Big\rangle =
\\ 
& \hskip 1cm   = e^{-\frac{\mu^2 (k_1^2+k_2^2)}{4}\log(\Lambda)} \, e^{- \frac{\mu}{\sqrt{2n}}\frac{1}{z_2 ^n} \vec{k}_{2} \cdot  \vec{\lambda}_{n}} e^{ \frac{\mu}{\sqrt{2n}}z_1 ^n \vec{k}_{1} \cdot  \vec{\lambda}_{n} ^*}e^{- \frac{\mu}{\sqrt{2n}}\frac{1}{z_1 ^n} \vec{k}_{1} \cdot \vec{\lambda}_{n}}e^{ \frac{\mu}{\sqrt{2n}}z_2 ^n \vec{k}_{2} \cdot \vec{ \lambda}_{n}^* }   e^{-\frac{\mu^2}{2} \vec{k}_1 \cdot \vec{k}_2 \sum_{n=1} ^{\Lambda} \frac{1}{n}\left( \frac{z_2}{z_1}\right)^n}~.
\nn
\end{align}
The argument in the exponential that contains the sum reads
\begin{equation}
	-\frac{\mu^2}{2} \vec{k}_1 \cdot \vec{k}_2 \sum_{m=1} ^{\Lambda} \frac{1}{m}\left( \frac{z_2}{z_1}\right)^m=\frac{\mu^2}{2} k_1 \tight\cdot k_2\, \prop~.
\end{equation}
where 
\begin{equation}\label{log}
\prop\equiv -\sum_{m=1} ^{\Lambda} \frac{1}{m}\left( \frac{z_2}{z_1}\right)^m \quad,\qquad \lim_{{\Lambda\to \infty}} \prop = \log \Big(1-\frac{z_2}{z_1}\Big)
\end{equation}
Therefore
\begin{align}
\label{2ptGreenFn}
&\bigg\langle \frac{1}{|\vec{x}-\vec{f}(z_1)|^2}\;\frac{1}{|\vec{y}-\vec{f}(z_2)|^2} \bigg\rangle = \frac{1}{(2\pi)^4}\int d^4 k_1 \int d^4 k_2\, \frac{e^{-i  \vec{k}_1 \cdot \vec{x} -i \vec{k}_2\cdot \vec{y}}}{|\vec{k}_1| ^2 |\vec{k}_2| ^2}
\left \langle e^{i \vec{k}_1 \cdot \vec{f}(z_1)} \; e^{i \vec{k}_2 \cdot \vec{f}(z_2)}\right\rangle
\nonumber\\[.2cm]	
&\hskip 2cm 
=\frac{1}{(2\pi)^4}\int d^4 k_1 \int d^4 k_2 \frac{e^{-i \vec{k}_1\cdot \vec{x} -i \vec{k}_2\cdot \vec{y}}}{|\vec{k}_1| ^2 |\vec{k}_2| ^2} \,e^{-\frac{\mu^2 (k_1^2+k_2^2)}{4}\log(\Lambda)} \,
\\[.2cm]
&\hskip4cm
e^{-\frac{\mu}{\sqrt{2n}}\frac{1}{z_2 ^n} \vec{k}_{2} \cdot \vec{ \lambda}_{n}+\frac{\mu}{\sqrt{2n}}z_1 ^n \vec{k}_{1}\cdot  \vec{\lambda}_{n} ^*- \frac{\mu}{\sqrt{2n}}\frac{1}{z_1 ^n} \vec{k}_{1} \cdot \vec{\lambda}_{n}+ \frac{\mu}{\sqrt{2n}}z_2 ^n \vec{k}_{2} \cdot \vec{\lambda}_{n}^* }
 e^{\frac{\mu^2 }{2} k_1 \cdot k_2 \prop}~.
 \nn
\end{align}
The real part of $\prop$ is bounded from below by $-\log(\Lambda)$, which takes place at coincident points - where the phases of eq. (\ref{log}) add up coherently. The eigenvalues of the matrix that characterizes the quadratic form in the argument of the exponential of (\ref{2ptGreenFn}) determine the convergence of the integral.  The quadratic form is  
\begin{equation}
	\frac{\mu^2}{4}
	\begin{pmatrix}
	-\log(\Lambda) & \prop \\
	\prop & -\log(\Lambda)	
	\end{pmatrix}
 ~,
\end{equation} 
and its eigenvalues are $-\frac{\mu^2}{4}(\log(\Lambda)\pm \prop)$, both having a negative real part; as a result, the integrals converge.
Note that if we drop the cutoff dependence, \eg\ by normal ordering, then the integrals do not converge.  Keeping it, however, won't affect the leading large N behavior we will find below; see Appendix~\ref{app:CompleteCalc} for details.
One can now carry out the integral over $\vec{k}_1$ in eq. (\ref{2ptGreenFn}) utilizing the formula (\ref{Fourierofx-2}).  
Defining
\begin{align}
	\Delta \vec{y} &\equiv \vec{y} -i\frac{\mu}{\sqrt{2n}}\left(  z_2 ^{-n} \vec{\lambda}_n-z_2 ^n \vec{\lambda}_n ^*\right),
\nn\\[.2cm]
	\Delta \vec{x} &\equiv \vec{x} -i\frac{\mu}{\sqrt {2n}}\left( z_1 ^{-n} \vec{\lambda}_n-z_1 ^n \vec{\lambda}_n ^* \right),
\end{align}
the integral in question reads
\begin{align}
\label{CorrelationIntegral0}
\left\langle \frac{1}{|\vec{x}-\vec{f}(z_1)|^2} \; \frac{1}{|\vec{y}-\vec{f}(z_2)|^2} \right\rangle 
= \frac{1}{(2\pi)^2} \int d^4 k_2 \, \frac{e^{ -i \vec{k}_2\cdot \Delta\vec{y}}}{ |\vec{k}_2| ^2} \frac{e^{-\frac{\mu^2 k_2 ^2}{4} \log(\Lambda)}}{\left|\Delta \vec{x}+i \frac{\mu^2}{2}\vec{k}_2\,\prop \right|^2}\left[1-e^{-\frac{\left|\Delta \vec{x}+i \frac{\mu^2}{2}\vec{k}_2\,\prop \right|^2}{\mu^2\log(\Lambda)}}\right]~.
\end{align}
We introduce notations for two of the terms that comprise the correlation function:
\begin{align}
\text{Corr}_L &\equiv \frac{1}{(2\pi)^2} \int d^4 k \, \frac{e^{ -i \vec{k}\cdot \Delta\vec{y}}}{ |\vec{k}| ^2} \frac{e^{-\frac{\mu^2 k ^2}{4} \log(\Lambda)}}{\left|\Delta \vec{x}+i \frac{\mu^2}{2}\vec{k}\,\prop \right|^2} 
\\[.3cm]
\text{Corr}_R &\equiv -\frac{1}{(2\pi)^2} \int d^4 k \, \frac{e^{ -i \vec{k}\cdot \Delta\vec{y}}}{ |\vec{k}| ^2} \frac{e^{-\frac{\mu^2 k ^2}{4} \log(\Lambda)}}{\left|\Delta \vec{x}+i \frac{\mu^2}{2}\vec{k}\,\prop \right|^2} e^{-\frac{\left|\Delta \vec{x}+i \frac{\mu^2}{2}\vec{k}\,\prop \right|^2}{\mu^2\log(\Lambda)}}.
\end{align}

%%%%%%%%%%%%%%%%%%%%%%
\subsection{Step 2: Evaluating integrals}
\label{subsec:Step2}

We begin by solving the following integral for any possible angle between $\Delta \vec{x}$ and $\Delta \vec{y}$:
\begin{align}
\text{Corr}_L  = \frac{1}{(2\pi)^2}\int \frac{d^4 k}{|k|^2} \frac{e^{-i\vec{k}\cdot \Delta \vec{y}-\frac{\mu^2 |\vec{k}| ^2}{4} \log(\Lambda)}}{\big| \Delta \vec{x} + i\frac{\mu^2}{2}\vec{k}\, \prop   \big|^2}~.
\end{align}
To this end, define:
\begin{equation}
	\vec{k}= \frac{\vec{\kappa}}{\Delta y} ~~,~~~~ \Delta \vec{x} = \Delta x ~ \hat{n}_x ~~,~~~~ \Delta \vec{y} = \Delta y ~ \hat{n}_y~.
\end{equation}
Suppose one places the unit vectors $\hat{n}_x$ and $\hat{n}_y$ on a common plane, where the angle between them is denoted by $\alpha$. We write
\begin{equation}
	\hat{n}_y = \hat{n}_x \cos\alpha + \hat{z} \sin\alpha~. 
\end{equation} 
Therefore,
\begin{align}
	\text{Corr}^{\vphantom{|}}_L = \frac{1}{(2\pi)^2 \Delta y ^2}\int \frac{d^4 \kappa}{|\kappa| ^2} \frac{e^{-i\vec{\kappa}\cdot \hat{n}_y-\frac{\mu^2 |\vec{\kappa}| ^2}{4|\Delta \vec{y}|^2} \log(\Lambda)}}{\big| \Delta \vec{x} + i\frac{\mu^2}{2\Delta y} \vec{\kappa}\,\prop  \big|^2}= -\frac{4}{(2\pi)^2 (\mu^2 \prop )^2 }\int \frac{d^4 \kappa}{|\kappa| ^2} \frac{e^{-i\vec{\kappa}\cdot \hat{n}_y}}{\left|\vec{\kappa}-2i\xi \hat{n}_x  \right|^2}e^{-\frac{\mu^2 |\vec{\kappa}|^2 \log(\Lambda)}{4|\Delta \vec{y}|^2}},\nonumber\\
\end{align}
where
\begin{equation}
	\xi \equiv \frac{\Delta x \Delta y}{\mu^2 \prop }~.
\end{equation}
Write
\begin{equation}
	d^4 \kappa = d\kappa_x \,\kappa^2 d\kappa \,d(\cos(\theta)) d\phi ~,~ |\kappa|^2 = \kappa_x ^2 + \kappa^2~,~
	\vec{\kappa} \cdot \hat{n}_y = \kappa \cos(\theta) \sin\alpha~. 
\end{equation}
We first integrate over the angles $0<\phi<2\pi$ and $0<\theta<\pi$:
\begin{align}\label{IntermediateStage}
	\text{Corr}^{\vphantom{|}}_L &= -\frac{4}{(2\pi)^2 (\mu^2 \prop )^2 }\int \frac{d\kappa_x \kappa^2 d\kappa d(\cos(\theta)) d\phi}{\kappa_x ^2+\kappa^2} \;\frac{e^{-i \kappa_x \cos\alpha-i\kappa \sin(\alpha )\cos(\theta)}}{\kappa_x ^2 + \kappa^2 -4i \kappa_x \xi  -4\xi^2}e^{-\frac{\mu^2 |\vec{\kappa}|^2 \log(\Lambda)}{4|\Delta \vec{y}|^2}} 
 \nonumber\\[.2cm]
	&=-\frac{4}{2\pi (\mu^2 \prop )^2 \sin\alpha }\int \frac{d\kappa_x \kappa d\kappa  ~2\sin(\kappa \sin\alpha)}{\kappa_x ^2+\kappa^2} \; \frac{e^{-i \kappa_x \cos\alpha -\frac{\mu^2 |\vec{\kappa}|^2 \log(\Lambda)}{4|\Delta \vec{y}|^2}}}{(\kappa_x -2i \xi - i \kappa)(\kappa_x -2i \xi + i \kappa)}~.
\end{align}
There are four different cases, corresponding to the signs of $\cos\alpha$ and $\text{Re}(\xi)$, all of which lead to the same result.  Let us work out the case $\cos\alpha>0$ and $\text{Re}(\xi)>0$,  
where one can close the $\kappa_x$ integration contour in the lower half plane. Two poles contribute:
\begin{equation}
	\kappa_x = -i\kappa \equiv\kappa_x^{(1)} ~~,~~~~ \kappa_x = -i\kappa +2i\xi \equiv\kappa_x^{(2)}~,
\end{equation}
where the latter contributes only if $2\text{Re}(\xi)<\kappa $. We write
\be
\text{Corr}_L = \text{Res}({\kappa_x^{(1)}}) + \text{Res}({\kappa_x^{(2)}}) 
\ee
Starting with the pole at $\kappa_x=-i\kappa$, one obtains
\begin{align}
	\text{Res}({\kappa_x^{(1)}}) & = -\frac{4}{2\pi (\mu^2 \prop )^2 \sin\alpha }\int \frac{ \kappa d\kappa  ~2\sin(\kappa \sin\alpha)}{-2i\kappa} (-2\pi i)\frac{e^{- \kappa \cos\alpha}}{(-2i \xi - 2i \kappa)(-2i \xi )}
 \nonumber\\
	&\hskip 2cm=\frac{1}{ \xi (\mu^2 \prop )^2 \sin\alpha }\int_0 ^{\infty}  d\kappa \sin(\kappa \sin\alpha) \frac{e^{- \kappa \cos\alpha}}{ \xi + \kappa}~.
 \end{align}
 Shifting the variable $\tilde{\kappa}\equiv \kappa+\xi$ and using the definition of the exponential integral function
\begin{equation}
E_1 (z) = \int_z ^{\infty} \frac{dt}{t} e^{-t}~,
\end{equation}
one has
 \begin{align}\label{FirstPole}
 	\text{Res}({\kappa_x^{(1)}}) & = \frac{e^{\xi \cos\alpha}}{ \xi (\mu^2 \prop )^2 2i\sin\alpha }\int_{\xi} ^{\infty} \frac{d\tilde{\kappa}}{\tilde{\kappa}}\left(e^{-i\xi \sin\alpha}e^{-\tilde{\kappa}(\cos(\alpha )-i\sin\alpha)}-e^{i\xi \sin\alpha}e^{-\tilde{\kappa}(\cos(\alpha )+i\sin\alpha)}\right)
  \nonumber\\[.3cm]
	&=\frac{1}{ \xi (\mu^2 \prop )^2 2i\sin\alpha }\left[ e^{\xi e^{-i\alpha}}E_1 \left(\xi e^{-i\alpha} \right)-e^{\xi e^{i\alpha}}E_1 \left(\xi e^{i\alpha} \right)\right]~.
\end{align}

In appendix \ref{app:CompleteCalc} we complete the calculation of the two-point function by computing the residue coming from the pole $\kappa_x ^{(2)}$ and also the contribution to the two-point function from the Gaussian. In that appendix we show that these terms are exponentially suppressed relative to the term in eq. (\ref{FirstPole}) by either $\frac{N}{n^2}$ or $\frac{N}{n^2 \log(\Lambda)}$ and thus we do not write them here.

While $\text{Corr}(z_1,z_2)$ is not real for general points $z_1,z_2$, when integrating them in the unit circle, one obtains a real answer if $\vec{x}=\vec{y}$, since one can for example sum pairs of the function $\text{Corr}$ at $(z_1,z_2)$ and $(z_2,z_1)$. 

%%%%%%%%%%%%%%%%%%%%%%%%%%%%%%%%%%%%
\subsection{Step 3:  Large \texorpdfstring{$N$}{} approximation}

We set $\Lambda = c N$ for some constant $c$, and take the large $N$ limit.
We will now show that for order one coherent state integer labels $n=O(1)$ and in all spatial regions except very close to the ring (how close to be specified below),  only the large $\xi$ limit of expression (\ref{FirstPole}) should be considered. 
We recall:
\begin{align}
	\xi &\equiv \frac{\Delta x \Delta y}{\mu^2 \log\left(1-e^{i(v_1-v_2)}\right)}
\nn\\[.2cm]
	\Delta \vec{x} &= \vec{x} - i \frac{\mu}{\sqrt{2n}}\left(\vec{\lambda}_n e^{inv_1}-\vec{\lambda}_n ^* e^{-inv_1}\right)
\nn\\[.2cm]
	\Delta \vec{y} &= \vec{y} - i \frac{\mu}{\sqrt{2n}}\left(\vec{\lambda}_n e^{inv_2}-\vec{\lambda}_n ^* e^{-inv_2}\right)
\\[.2cm]
	\vec{\lambda}_n &= \lambda (1,i,0,0)~~,~~~~
	a = \sqrt{\frac{2}{n}}\mu\lambda~~,~~~~
 \mu= \frac{g_s (\alpha')^2}{R_y  \sqrt{V_4}}~.
\nn
\end{align}
\begin{equation}
	a= \frac{\mu \sqrt{n_1 n_5}}{n}~. 
\end{equation} 
Therefore,
\begin{equation}
	\Delta \vec{x} = \vec{x} +a(\sin(nv_1),\cos(nv_1),0,0) ~~,~~~~ \Delta \vec{y} = \vec{y}+a (\sin(nv_2),\cos(nv_2),0,0) ~. 
\end{equation}
Further defining
\begin{equation}
	N = n_1 n_5  ~~,~~~~
	\tilde{x} = \frac{\vec{x}}{a} ~~,~~~~ \tilde{y} = \frac{\vec{y}}{a}~,
\end{equation}
it follows that
\begin{equation}
	\xi =\frac{a^2 \Delta \tilde{x} \Delta \tilde{y}}{\mu^2 \log\left(1-e^{i(v_1-v_2)}\right)} =\frac{N}{n^2}\frac{\Delta \tilde{x} \Delta \tilde{y}}{\log\left(1-e^{i(v_1-v_2)}\right)} ~.
\end{equation}
For large arguments, the exponential integral can be approximated by \cite{AbraSteg72}
\begin{equation}
	E_1 (z) \sim \frac{e^{-z}}{z}\left( 1-\frac{1}{z}+\frac{2}{z^2}+O\left(\frac{1}{z^3}\right)\right) ~~,~~~~ |\text{arg}(z)|< \frac{3\pi}{2}~.
\end{equation}
Then for $|\xi| \gg 1$ the result (\ref{FirstPole}) approximates to
\begin{align}
\label{corrapprox}
\text{Corr}^{\vphantom{|}}_L &= \frac{1}{\mu^4 \prop ^2 }\bigg(\frac{1}{\xi^2}-\frac{2\cos\alpha}{\xi^3}+\ldots \bigg)
= \text{Corr}_L^{(1)} + \text{Corr}_L^{(2)} + \ldots
\end{align}
Since $\xi \propto \frac{N}{n^2}$, the higher order terms are suppressed by higher powers of $\xi$ are negligible in the regime $1\ll \frac{N}{n^2}$. We thus drop them in what follows and focus on the two terms in the top line of~eq. (\ref{corrapprox}).

The first term $\text{Corr}_L^{(1)}$ in the large-argument approximation corresponds the factorized expression:
\begin{equation}
	\text{Corr}_L^{(1)} = \frac{1}{\Delta x ^2 \Delta y^2}~.
\end{equation}
The integrals over $v_1,v_2$ yield the product of one-point functions.
Recalling the review in subsection~\ref{subsec:HarmFnsReview}, and specifically eqs.~\eqref{Result4DGreensFn2}, \eqref{zint-ring}, we have
\begin{equation}
\frac{1}{(2\pi)^2}\int_0 ^{2\pi}\! dv_1\int_0 ^{2\pi}\! dv_2\, \text{Corr}_L^{(1)} =
	H_5 (\tilde{x})H_5 (\tilde{y}) = \frac{Q_5 ^2}{a^4 \sqrt{(\tilde{x}^2+1)^2 - 4(\tilde{x}_1 ^2 + \tilde{x}_2^2)^2}\sqrt{(\tilde{y}^2+1)^2 - 4(\tilde{y}_1 ^2 + \tilde{y}_2^2)^2}}~.
\end{equation}
In the last equation, the variables $v_1,v_2$ are rescaled by $\frac{2\pi}{L}$ relative to the ones in eq.~(\ref{greensfn}).
In computing the relative fluctuations, we subtract the product of the one-point function from the two-point function, and normalize by the product of the one-point functions, 
\be
\label{relfluctsdef}
\text{RelFlucts}(\cO(\vec{x},\vec{y})) = \frac{\vev{\cO(\tilde x)\cO(\tilde y)}-\vev{\cO(\tilde x)}\vev{\cO(\tilde y)}}{\vev{\cO(\tilde x)}\vev{\cO(\tilde y)}}
\ee
and so the leading term in the fluctuations comes from the next term, $\text{Corr}_L^{(2)}$.

\begin{figure}[ht]
\vskip .5cm
\centering
  \begin{subfigure}[b]{0.43\textwidth}
  \hskip 0cm
    \includegraphics[width=\textwidth]{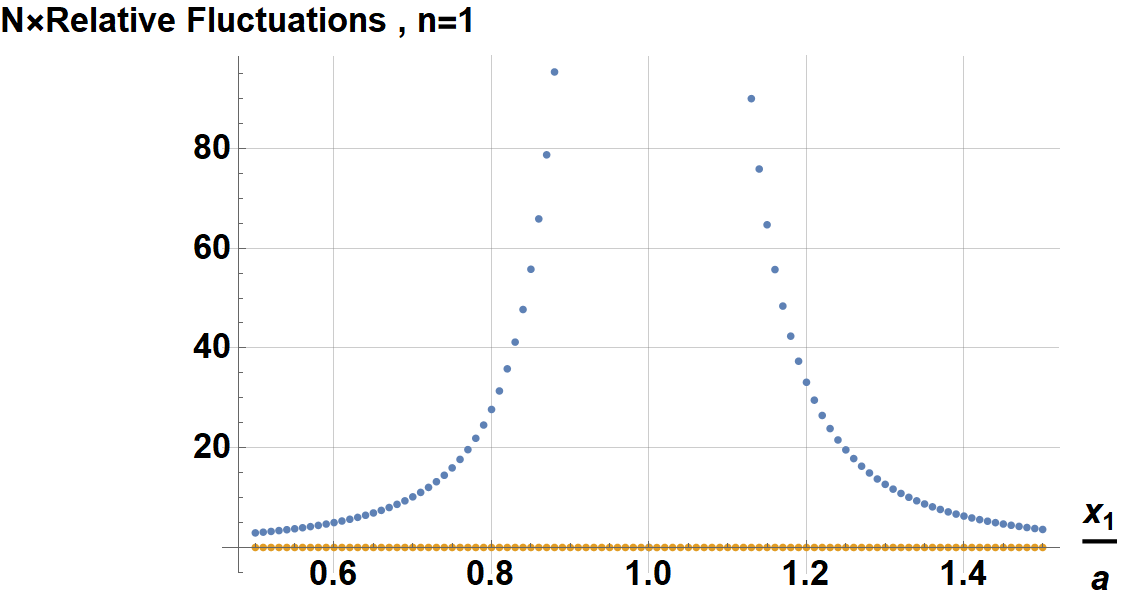}
    \caption{ }
    \label{fig:3alabel}
  \end{subfigure}
\qquad\qquad
  \begin{subfigure}[b]{0.43\textwidth}
  \vskip -2cm
    \includegraphics[width=\textwidth]{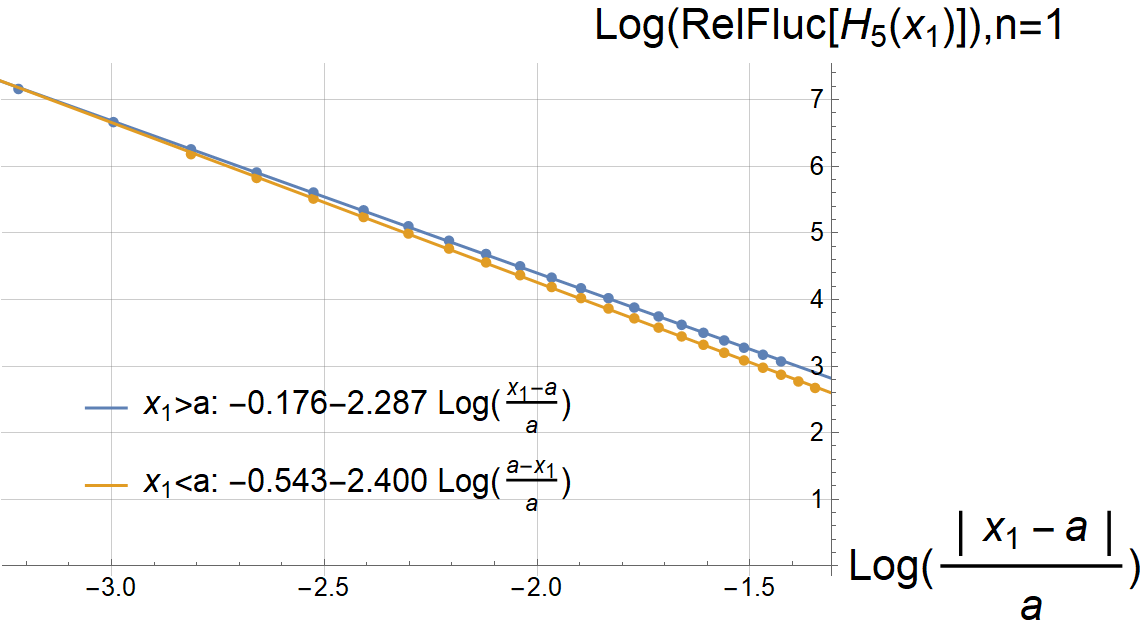}
    \caption{ }
    \label{fig:3blabel}
  \end{subfigure}
\caption{\it 
a) The leading large $N$ contribution to the relative fluctuations of the metric components in transverse space for $n=1$. As a function of the position along the plane of the ring, the blue points show the real parts and the orange points the imaginary parts (the latter should vanish; that they do is a check on the numerical accuracy).  
b) Linear fit of the log of the relative fluctuations of $H_5$ versus the log of the fractional deviation from the ring radius $a$, for 20 points each inside and outside the ring, along the $x_1$ axis. The intercept in the fit for points with $x_1>a$ is $-0.17\pm 0.01$ and the slope is $-2.287\pm 0.004$. The intercept in the fit for points with $x_1<a$ is $-0.54\pm 0.01$ and the slope is $-2.400\pm 0.006$. 
}
\label{fig:StatPhaseApproxPlaneRing}
\end{figure}

Moving on to this term, we have
\begin{align}
\label{2ndTerm}
	& \text{Corr}_L^{(2)} = -\frac{2\cos\alpha}{(\mu^2 \prop )^2 \xi^3} = -\frac{2\Delta \tilde{\vec{x}}\cdot \Delta \tilde{\vec{y}}}{\Delta \tilde{x}^4 \Delta \tilde{y}^4} \frac{ n^2 \log\left(1-e^{i(v_1-v_2)}\right)}{a^4 N}~.
\end{align}
Thus the leading contribution to the relative fluctuations is equal to
\begin{align}
\label{Fluctuations2}
\text{RelFlucts}(H_5(\vec{x},\vec{y})) &\approx -2\frac{n^2}{N}\sqrt{(\tilde{x}^2+1)^2 - 4(\tilde{x}_1 ^2 + \tilde{x}_2^2)^2}\sqrt{(\tilde{y}^2+1)^2 - 4(\tilde{y}_1 ^2 + \tilde{y}_2^2)^2} \nonumber\\
	&\hskip 2cm\times \frac{1}{(2\pi)^2}\int_0 ^{2\pi}  dv_1 \int_0 ^{2\pi} dv_2 \frac{\Delta \tilde{\vec{x}}\cdot \Delta \tilde{\vec{y}}}{\Delta \tilde{x}^4 \Delta \tilde{y}^4}  \log\left(1-e^{i(v_1-v_2)}\right)~.
\end{align}
For $\vec{x}=\vec{y}$, $x_2\tight=x_3\tight=x_4\tight=0$, $n=1$ and in a direction $\frac{1}{2}<{x_1}/{a}<\frac{3}{2}$ in the plane of the ring, a numerical integration of $ N\times$(eq.\;\ref{Fluctuations2}) using Mathematica is depicted in figure~\ref{fig:StatPhaseApproxPlaneRing}. 

We see that the behavior of the relative fluctuations near the ring is well-approximated by a power law.  
For order one values of $n$ and $x_1\gtrsim a$, the exponent of the power is approximately $-2.287$ independent of~$n$.
Furthermore, for $n=1,2,3,4,5$ we find that, approaching the ring from outside $x_1\gtrsim a$, the overall coefficient of a power law fit to the relative fluctuations scales linearly with $n$ with a slope $0.838\pm 0.0001$ and an intercept that is zero to a precision $4\times 10^{-4}$. In other words,
\begin{align}
\label{relfluctsfit}
    \text{RelFlucts}(H_5(x_1))\approx \frac{0.838\,n}{N}\frac{1}{(\tilde{x}_1 - 1)^{2.287}}~.
\end{align}
The fluctuations are of order one when
\begin{equation}
 \frac{x_1-a}{a} \approx 
\left(\frac{0.838\,n}{N}\right)^{\frac{1}{2.287}}.
\end{equation}
The invariant distance between the point $x_1=a+a\left(\frac{n}{N}\right)^{\frac{1}{2.287}}$ and the point $x_1=a+2a\left(\frac{n}{N}\right)^{\frac{1}{2.287}}$, according to the Lunin-Mathur line element,  is proportional to
\begin{equation}\label{Ddistance}
D \propto  \sqrt{n_5 \alpha'} \left(\frac{n}{N}\right)^{\frac{1}{4.57}}  ~.
\end{equation}
This scale is slightly bigger than the 6D Planck scale~\eqref{PlanckScale6D}.  If the power law in~\eqref{Ddistance} had been 1/4 instead of 1/4.57, the fluctuations would be growing large at a 6d Planck distance from the source.  Instead, because of the power law resulting from the numerical analysis of~\eqref{Fluctuations2}, the fluctuations become of order one at a scale bigger than the 6d Planck scale by a factor $N^{0.03}$.  This is still much smaller than the blob size $L$ (eq.~\eqref{blobsize}) of the ensemble geometry~\eqref{ensgeom} by a factor $N^{0.47}$.

Repeating the analysis for points in the plane of the ring but slightly inside it ($x_1\lesssim a$), one finds a similar power law behavior, but with a power $2.400\pm 0.006$ instead of $2.287\pm 0.004$.
\begin{align}
    \text{RelFlucts}\big(H_5(x_1)\big)\approx \frac{0.581 n}{N}\frac{1}{( 1-\tilde{x}_1)^{2.400}}~.
\end{align}
There is again a region of space inside and close to the ring where the metric fluctuations are of order one. This region is bigger than the 6D Planck length by $N^{0.04}$. This power law fit is valid near the radius of the ring; along the $x_1$ axis and near the center of the ring, the relative fluctuations approach the constant $\frac{n}{N}$.

Next, we consider the region transverse to the plane of the ring, along the axis passing through its center $x_{12}=0$. Then expression (\ref{Fluctuations2}) admits an analytical form, 
\begin{equation}
	\text{RelFlucts}\big(H_5(x_{34})\big)=-\frac{n^2}{N}\big(\tilde{x}_{34}^2+1\big)^2 \bigg(-\frac{1}{n (1+\tilde{x}_{34}^2)^4}\bigg)=\frac{n}{N} \frac{1}{(1+\tilde{x}_{34}^2)^2}~.
\end{equation}
This result shows that along the symmetry axis of the ring, the relative quantum fluctuations of the metric components are everywhere small when $n \ll N$.

%%%%%%%%%%%%%%%%%%%%%%%%%%%%%%%%%%%%%
%%%%%%%%%%%%%%%%%%%%%%%%%%%%%%%%%%%%%

\section{Discussion}
\label{sec:disc}

The geometry of 1/2-BPS NS5-F1 bound states~\eqref{LMgeom} is parametrized by harmonic functions \eqref{greensfn} specified by the fivebrane source configuration.
A typical 1/2-BPS NS5-F1 bound state in the ensemble of such states at large fixed charges $n_1,n_5$ is well-approximated by an ``ensemble geometry'', in which the harmonic functions are averaged over source configurations, with the result~\eqref{ensgeom}~\rcite{Alday:2006nd}. 

One has to apply a T-duality transformation on the spatial circle $\bS^1_y$ to describe the system near the source~\rcite{Martinec:1999gw,Chen:2014loa,PaperA}. The geometry exhibits a large redshift near the core where the fuzzy bound state is quasi-localized, with a spherical blob structure of size set by the AdS scale $\ell_{AdS}=\sqrt{n_5 \alpha'}$. The T-dual circle $\tilde\bS^1_y$ has a large circumference near the source and the $\bS^3$ factor of the geometry collapses smoothly at the origin, where the magnetic NS-NS flux no longer supports it at a fixed positive radius.

There are several possible failure modes for the use of effective field theory to describe these typical two-charge backgrounds:
\begin{enumerate}
\item
Large curvature, so that higher-derivative corrections to the action and equations of motion become important.  We have seen in~\rcite{PaperA} that in the generic 1/2-BPS state, this issue does not arise if one works in the appropriate (NS5-P) duality frame near the source.
\item
Fivebrane intersections, so that non-abelian fivebrane dynamics arises at the intersection, and perturbative string theory breaks down.  It was shown in~\rcite{PaperA} that such intersections are statistically rare in the ensemble of 1/2-BPS states.
\item
Large quantum fluctuations in the geometry.  We have seen here that the fluctuations in the harmonic function $H_5$ in the geometry~\eqref{LMgeom}, \eqref{circharmfns} are small until one gets very close to the source~-- somewhat larger than the 6d Planck scale, but parametrically much smaller than the AdS scale $\ell_{AdS}$ by a power of $N$.  Given that the typical coherent state source is well separated from itself as it wanders about, we expect that quantum fluctuations of the geometry in such a state are also small.  
\end{enumerate}
Regarding this last point, a second source of fluctuations arises when performing a statistical average over the phase space of classical solutions, which is closely related to an integration over coherent state parameters in the quantum theory (\cf\ the analysis in~\rcite{Raju:2018xue,Alday:2006nd,PaperA}).
The classical version of these statistical fluctuations  consists of an integral of correlation observables such as~\eqref{2ptGreenfn0} over the classical phase space.  In a correlator of exponentials between coherent states such as~\eqref{exptl2ptfn}, the classical correlator arises when one considers only the action of the annihilation operators on the coherent state, which yields the coherent state parameter.  The commutators of the mode operators among themselves are absent classically.%
\footnote{Indeed, the analysis of~\rcite{Raju:2018xue} dropped such normal-ordering terms, so they are working with the classical ensemble average.  Had they kept the normal-ordering terms that they do consider, the resulting integrals would have diverged in a fixed coherent state (see the discussion following~\eqref{2ptGreenFn} above).  These divergences go away if one averages over coherent state parameters first, and then does the remaining integrals.}
This leads to a classical correlator which factorizes, and to vanishing relative fluctuations~\eqref{relfluctsdef} in a given classical state.  

In the quantum theory, these commutator terms are of course present~-- they lead to all the factors gaussian in momenta in the correlator~\eqref{2ptGreenFn}~-- but their effects are $1/N$ suppressed in a fixed coherent state, as the result~\eqref{relfluctsfit} shows. 

The ensemble average in the classical theory leads to non-vanishing relative fluctuations of a different type. These arise from averaging the joint expectation value of classical observables over the coherent state parameters which parametrize the classical phase space (\eg\ observables such as the analogue of eq.~\eqref{2ptGreenFn} in a general coherent state, with operator ordering factors dropped).  This leads to a much larger (in fact independent of $N$ at leading order in the $1/N$ expansion) classical statistical fluctuation of quantities such as the two-point correlator~\eqref{2ptGreenfn0}.
These classical statistical fluctuations are the effect seen in the calculation of~\cite{Raju:2018xue}.  The quantum contributions to the relative fluctuations are small in each coherent state, and remain small upon averaging over coherent states.

We thus conclude that the large fluctuations observed in~\rcite{Raju:2018xue} in the ensemble average of two-charge BPS microstates, are statistical fluctuations of the ensemble rather than quantum fluctuations of the sources or of the geometry surrounding them, the latter being small in any particular member of the ensemble.  We expect this picture to hold true also when the underlying states preserve fewer supersymmetry generators, \eg\ for states obtained from two-charge states by spectral flow~\rcite{Giusto:2004id,Jejjala:2005yu}, for which there is a valid perturbative string description~\rcite{Martinec:2017ztd}; or for NS5-F1-P quarter-BPS superstrata~\rcite{Bena:2015bea,Bena:2017xbt}, which are smooth deformations of these backgrounds.

The generic 1/2-BPS fivebrane source is a non-selfintersecting, classical configuration spread out over a region of proper size much larger than the Planck and string scales, whose proper size is of order the AdS scale or more in the radial direction (see~\rcite{PaperA}, eq. 2.82), and of order the AdS scale in the angular directions.  There are no apparent sources of breakdown of classical string theory in the description of an individual such microstate.  These objects are thus much better thought of as fivebrane stars than as small black holes.

%%%%%%%%%%%%%

\vspace{1cm}

\section*{Acknowledgements}
%%%%%%%%%%%%%

%
%
The work of EJM is supported in part by DOE grant DE-SC0009924. The work of YZ is supported by the Blavatnik fellowship, and was supported by the Adams fellowship and the German Research Foundation through a
German-Israeli Project Cooperation (DIP) grant ``Holography and the Swampland''. YZ thanks the University of Chicago and IPhT Saclay for their hospitality during his visits.

%%%%%%%%%%%%%%%%%%%%%%%%%%%%%%%%%%%%%
%%%%%%%%%%%%%%%%%%%%%%%%%%%%%%%%%%%%%

%\vskip 1cm
%\appendix

\appendix

\section{Completing the two-point function calculation}

\label{app:CompleteCalc}

The goal of this appendix is to arrive at an expression for the two-point function of the four-dimensional Green's functions, and show that a particular contribution to it dominates at large $N$. The expression we start with is
\begin{align}
\label{CorrelationIntegral11}
\left\langle \frac{1}{|\vec{x}-\vec{f}(z_1)|^2} \; \frac{1}{|\vec{y}-\vec{f}(z_2)|^2} \right\rangle 
= \frac{1}{(2\pi)^2} \int d^4 k \, \frac{e^{ -i \vec{k}\cdot \Delta\vec{y}}}{ |\vec{k}| ^2} \frac{e^{-\frac{\mu^2 k ^2)}{4} \log(\Lambda)}}{\left|\Delta \vec{x}+i \frac{\mu^2}{2}\vec{k}\,\prop \right|^2}\left[1-e^{-\frac{\left|\Delta \vec{x}+i \frac{\mu^2}{2}\vec{k}\,\prop \right|^2}{\mu^2\log(\Lambda)}}\right]~.
\end{align}
In subsection \ref{subsec:Step2} we showed that the term coming from the $1$ in the square brackets can be written as (\ref{IntermediateStage})
\begin{equation}\label{Integralskappaxkappa}
    -\frac{4}{2\pi (\mu^2 \prop )^2 \sin\alpha }\int \frac{d\kappa_x \kappa d\kappa  ~2\sin(\kappa \sin\alpha)}{\kappa_x ^2+\kappa^2} \; \frac{e^{-i \kappa_x \cos\alpha -\frac{\mu^2 |\vec{\kappa}|^2 \log(\Lambda)}{4|\Delta \vec{y}|^2}}}{(\kappa_x -2i \xi - i \kappa)(\kappa_x -2i \xi + i \kappa)}=\text{Res}({\kappa_x^{(1)}})+\text{Res}({\kappa_x^{(2)}}).
\end{equation}
This receives a contribution from a pole at $\kappa_x ^{(1)}=-i\kappa$ which we computed to be~\eqref{FirstPole}, which we reproduce here for convenience:
\begin{align}
 	\text{Res}({\kappa_x^{(1)}}) =\frac{1}{ \xi (\mu^2 \prop )^2 2i\sin\alpha }\left[ e^{\xi e^{-i\alpha}}E_1 \left(\xi e^{-i\alpha} \right)-e^{\xi e^{i\alpha}}E_1 \left(\xi e^{i\alpha} \right)\right]~.
\end{align}
It is useful to write this result in the following form (assuming for simplicity that $\sin(\alpha)>0$):
\begin{align}
&\frac{1}{  \mu^2 \prop  2i |\Delta \vec{x}\wedge \Delta \vec{y}| }\left[ e^{\frac{1}{\mu^2 \prop}\left(\Delta \vec{x} \cdot \Delta \vec{y}- i |\Delta \vec{x}\times \Delta \vec{y}|\right)}E_1 \left(\frac{1}{\mu^2 \prop}\left(\Delta \vec{x} \cdot \Delta \vec{y}- i |\Delta \vec{x}\wedge \Delta \vec{y}|\right) \right)\right. \nonumber\\
	&\left. -e^{\frac{1}{\mu^2 \prop}\left(\Delta \vec{x} \cdot \Delta \vec{y}+ i |\Delta \vec{x}\wedge \Delta \vec{y}|\right)}E_1 \left(\frac{1}{\mu^2 \prop}\left(\Delta \vec{x} \cdot \Delta \vec{y}+ i |\Delta \vec{x}\wedge \Delta \vec{y}|\right)\right)\right].
\end{align} 
Here we have defined
\begin{equation}
    |\Delta \vec{x} \wedge \Delta \vec{y}|\equiv |\Delta \vec{x}| |\Delta \vec{y}|-\Delta\vec{x}\cdot \Delta \vec{y}.
\end{equation}

A second contribution arises from a pole at $\kappa_x ^{(2)} = -i\kappa+2i \xi$, which is inside the complex contour if $2\text{Re}(\xi)<\kappa$, can be calculated utilizing the residue theorem for the integral over $\kappa_x$ in (\ref{Integralskappaxkappa}):
\begin{align}\label{2}
	&\text{Res}(\kappa_x ^{(2)})= -\frac{4e^{2\xi \cos\alpha}}{2\pi (\mu^2 \prop )^2 \sin\alpha }\int_{2\text{Re}(\xi)} ^{\infty} \frac{\kappa d\kappa  ~2\sin(\kappa \sin\alpha)}{-4\xi^2 +4\xi \kappa } \frac{e^{- \kappa \cos\alpha} (-2\pi i)}{( -2 i \kappa)} e^{-\frac{\mu^2 \log(\Lambda)}{4|\Delta \vec{y}|^2} (-4\xi^2 +4\xi \kappa )}
	\nonumber\\[.2cm]
	&\hskip 1cm=-\frac{e^{2\xi \cos\alpha}}{\xi (\mu^2 \prop )^2 \sin\alpha }\int_{2\text{Re}(\xi)} ^{\infty} \frac{d\kappa  ~\sin(\kappa \sin\alpha)}{-\xi + \kappa } e^{- \kappa \cos\alpha}e^{-\frac{\mu^2 \xi \log(\Lambda)}{|\Delta \vec{y}|^2} (-\xi +\kappa )}=[\tilde{\kappa}=\kappa-\xi]
	\nonumber\\[.2cm]
	&\hskip 1cm=-\frac{e^{\xi \cos\alpha}}{\xi (\mu^2 \prop )^2~2i \sin\alpha }\int_{\xi^*} ^{\infty} \frac{d\tilde{\kappa} }{\tilde{\kappa} } e^{- \tilde{\kappa }\cos\alpha}\left(e^{i\xi \sin\alpha}e^{i\tilde{\kappa}\sin\alpha}-e^{-i\xi \sin\alpha}e^{-i\tilde{\kappa}\sin\alpha}\right)e^{-\frac{\mu^2 \xi \log(\Lambda)}{|\Delta \vec{y}|^2} \tilde{\kappa}}
	\nn\\[.2cm]
	&\hskip 1cm=-\frac{1}{\xi (\mu^2 \prop )^2~2i \sin\alpha } \left[e^{\xi e^{i\alpha}} E_1 \left(\xi^* e^{-i\alpha}+\frac{\mu^2 |\xi|^2 \log(\Lambda)}{|\Delta \vec{y}|^2} \right)-e^{\xi e^{-i\alpha}} E_1 \left(\xi^* e^{i\alpha}+\frac{\mu^2 |\xi|^2 \log(\Lambda)}{|\Delta \vec{y}|^2}\right)\right]~.
\end{align}
Next, we move to
\begin{align}
	&\text{Corr}_R \equiv -\frac{1}{(2\pi)^2}\int \frac{d^4 k}{|\vec{k}|^2} e^{-i\vec{k} \cdot \Delta \vec{y}} e^{-\frac{k ^2 \mu^2}{4} \log(\Lambda)}\frac{1}{|\Delta\vec{x}+i\frac{\mu^2}{2} \vec{k} \prop  |^2}e^{-\frac{\Big|\Delta\vec{x}+i\frac{\mu^2}{2} \vec{k} \prop  \Big|^2}{\mu^2 \log(\Lambda)}}=\nonumber\\
	&-\frac{1}{(2\pi)^2}e^{-\frac{|\Delta \vec{x}|^2}{\mu^2 \log(\Lambda)}}\int \frac{d^4 k}{|\vec{k}|^2} e^{-i\vec{k} \cdot \left(\Delta \vec{y}+\mu^2 \prop \Delta \vec{x}\right)} e^{-\frac{k ^2 \mu^2}{4}\left(\log(\Lambda) -\frac{\prop^2}{\log(\Lambda)}\right)}\frac{1}{|\Delta\vec{x}+i\frac{\mu^2}{2} \vec{k} \prop  |^2}.
\end{align}
Relative to the $\text{Corr}_L$, one sees that 
\begin{equation}
\Delta \vec{y} \longrightarrow \Delta \vec{y} +  \frac{\prop}{\log(\Lambda)} \Delta \vec{x}
\quad,\qquad
\log(\Lambda) \longrightarrow \log(\Lambda) - \frac{\prop^2}{\log(\Lambda)}	
\end{equation}
 and an overall Gaussian factor $-e^{-\frac{|\Delta \vec{x}|^2}{\mu^2 \log(\Lambda)}}$ exists for $\text{Corr}_R$ but not for $\text{Corr}_L$. 
Then coming back to evaluating $\text{Corr}_R$, two contributions from the two poles emanate. Starting with the pole at $\kappa_x ^{(1)} =-i\kappa$, the associated contribution is given by
\begin{align}\label{3}
		&\text{Res}(\kappa_x ^{(1)})_R= -\frac{1}{ \xi (\mu^2 \prop )^2 2i\sin\alpha }\left[ e^{\xi e^{-i\alpha}}E_1 \left(\xi e^{-i\alpha}+\frac{|\Delta \vec{x}|^2}{\mu^2\log(\Lambda)} \right)-e^{\xi e^{i\alpha}}E_1 \left(\xi e^{i\alpha}+\frac{|\Delta \vec{x}|^2}{\mu^2\log(\Lambda)}  \right)\right].
\end{align}
The contribution from the pole at $\kappa_x ^{(2)} = -i\kappa+2i\xi$ is given by
\begin{align}\label{4}
	&\text{Res}(\kappa_x ^{(2)})_R= \frac{1}{\xi (\mu^2 \prop )^2~2i \sin\alpha }\left[e^{\xi e^{i\alpha}} E_1 \left(\xi^* e^{-i\alpha}+\frac{|\Delta \vec{x}|^2}{\mu^2\log(\Lambda)}+\frac{\mu^2 |\xi|^2 \left(\log(\Lambda)-\frac{\prop^2}{\log(\Lambda)}\right)}{|\Delta \vec{y}|^2} \right) \right.\nonumber\\
	&-\left. e^{\xi e^{-i\alpha}} E_1 \left(\xi^* e^{i\alpha}+\frac{|\Delta \vec{x}|^2}{\mu^2\log(\Lambda)}+\frac{\mu^2 |\xi|^2 \left(\log(\Lambda)-\frac{\prop^2}{\log(\Lambda)}\right)}{|\Delta \vec{y}|^2}\right)\right].
\end{align}
The full result is now
\begin{align}
	&\Big\langle  \frac{1}{|\vec{x}-\vec{f}(z_1)|^2 } \frac{1}{|\vec{y}-\vec{f}(z_2)|^2 } \Big\rangle =\frac{1}{ \xi (\mu^2 \prop )^2 2i\sin\alpha }\times  \nonumber\\ 
	&\left[ e^{\xi e^{-i\alpha}}E_1 \left(\xi e^{-i\alpha} \right)-e^{\xi e^{i\alpha}}E_1 \left(\xi e^{i\alpha} \right)\right.\nonumber\\
	&\left. -e^{\xi e^{i\alpha}} E_1 \left(\xi^* e^{-i\alpha}+\frac{|\Delta \vec{x}|^2 \log(\Lambda)}{\mu^2 |\prop|^2} \right)+e^{\xi e^{-i\alpha}} E_1 \left(\xi^* e^{i\alpha}+\frac{|\Delta \vec{x}|^2 \log(\Lambda)}{\mu^2 |\prop|^2}\right) \right.~\nonumber\\
     &\left. -e^{\xi e^{-i\alpha}}E_1 \left(\xi e^{-i\alpha}+\frac{|\Delta \vec{x}|^2}{\mu^2\log(\Lambda)} \right)+e^{\xi e^{i\alpha}}E_1 \left(\xi e^{i\alpha}+\frac{|\Delta \vec{x}|^2}{\mu^2\log(\Lambda)}  \right) \right. \nonumber\\
	 &\left. +e^{\xi e^{i\alpha}} E_1 \left(\xi^* e^{-i\alpha}+\frac{|\Delta \vec{x}|^2 \log(\Lambda)}{\mu^2 |\prop|^2}+\frac{|\Delta \vec{x}|^2 \left(1-e^{2i\text{arg}(\prop)}\right)}{\mu^2\log(\Lambda)} \right)  \right. \nonumber\\
	 &-\left. e^{\xi e^{-i\alpha}} E_1 \left(\xi^* e^{i\alpha}+\frac{|\Delta \vec{x}|^2 \log(\Lambda)}{\mu^2 |\prop|^2}+\frac{|\Delta \vec{x}|^2 \left(1-e^{2i\text{arg}(\prop)}\right)}{\mu^2\log(\Lambda)}\right) \right].
\end{align}
Let us perform a large argument expansion using \cite{AbraSteg72}
\begin{equation}
	E_1 (|z|\gg1) \sim \frac{e^{-z}}{z}\left(1-\frac{1}{z}+O\left(\frac{1}{z^2}\right)\right).
\end{equation}
Keeping the first few terms in that expansion, one has
\begin{align}\label{FinalAnswer}
	&\Big\langle \vec{\alpha}_n \Big| \frac{1}{|\vec{x}-\vec{f}(z_1)|^2 } \frac{1}{|\vec{y}-\vec{f}(z_2)|^2 }\Big|\vec{\alpha}_n \Big\rangle \approx \frac{1}{ \xi (\mu^2 \prop )^2 2i\sin\alpha }\times  \nonumber\\ 
	&\left[ 2i \sin(\alpha) \frac{1}{\xi} -\frac{2i \sin(2\alpha)}{\xi^2}\right.\nonumber\\
	&\left. -\frac{e^{\xi e^{i\alpha}-\xi^* e^{-i\alpha}}}{\xi^* e^{-i\alpha}+\frac{|\Delta \vec{x}|^2 \log(\Lambda)}{\mu^2 |\prop|^2}} e^{-\frac{|\Delta \vec{x}|^2 \log(\Lambda)}{\mu^2 |\prop|^2} }+\frac{e^{\xi e^{-i\alpha}-\xi^* e^{i\alpha}} }{\xi^* e^{i\alpha}+\frac{|\Delta \vec{x}|^2 \log(\Lambda)}{\mu^2 |\prop|^2}}e^{-\frac{|\Delta \vec{x}|^2 \log(\Lambda)}{\mu^2 |\prop|^2}} \right.~\nonumber\\
	& \left.  -\frac{e^{-\frac{|\Delta \vec{x}|^2}{\mu^2 \log(\Lambda)}}}{\zeta e^{-i\alpha} + \frac{|\Delta \vec{x}|^2}{\mu^2 \log(\Lambda)}}+\frac{e^{-\frac{|\Delta \vec{x}|^2}{\mu^2 \log(\Lambda)}}}{\zeta e^{i\alpha} + \frac{|\Delta \vec{x}|^2}{\mu^2 \log(\Lambda)}}\right. \nonumber\\
	&\left. +\frac{e^{\xi e^{i\alpha}-\xi^* e^{-i\alpha}}}{\xi^* e^{-i\alpha}+\frac{|\Delta \vec{x}|^2 \log(\Lambda)}{\mu^2 |\prop|^2}+\frac{|\Delta \vec{x}|^2 (1-e^{2i \text{arg}(\prop)})}{\mu^2 \log(\Lambda)}} e^{-\frac{|\Delta \vec{x}|^2}{\mu^2}\left[ \frac{\log(\Lambda)}{|\prop|^2}+\frac{1-e^{2i \text{arg}(\prop)}}{\log(\Lambda)}\right]}\right. \nonumber\\ 
	&\left. - \frac{e^{\xi e^{-i\alpha}-\xi^* e^{i\alpha}}}{\xi^* e^{-i\alpha}+\frac{|\Delta \vec{x}|^2 \log(\Lambda)}{\mu^2 |\prop|^2}+\frac{|\Delta \vec{x}|^2 (1-e^{2i \text{arg}(\prop)})}{\mu^2 \log(\Lambda)}} e^{-\frac{|\Delta \vec{x}|^2}{\mu^2}\left[ \frac{\log(\Lambda)}{|\prop|^2}+\frac{1-e^{2i \text{arg}(\prop)}}{\log(\Lambda)}\right]}\right].
\end{align}
Now, in the scaling limit where $N\to \infty$, $\Lambda = c N$ with $n$, $a$ and $c$ fixed, then the third, fifth and sixth lines of expression (\ref{FinalAnswer}) are suppressed relative to the second line by a factor of $e^{-\frac{|\Delta \vec{x}|^2 \log(\Lambda)}{\mu^2 |\prop|^2}}=e^{-\Delta \tilde{x}^2 \frac{N\log(c N)}{n^2 |\prop|^2}}$.
The fourth line of expression (\ref{FinalAnswer}) is suppressed by 
$e^{-\frac{|\Delta \vec{x}|^2}{\mu^2 \log(\Lambda)}}=e^{-\Delta \tilde{x}^2 \frac{N}{n^2 \log(c N)}}$ relative to the second line. Consequently, it is possible to neglect them and keep the second line as a reliable large $N$ approximation to the two-point function.

\newpage
%\vskip 2cm

\bibliographystyle{JHEP}      

\bibliography{fivebranes}

%%%%%%%%%%%%%%%%%%%%%%%%%%%%%%%%%%%%%%
%%%%%%%%%%%%%%%%%%%%%%%%%%%%%%%%%%%%%%

\end{document}

%%%%%%%%%%%%%%%%%%%%%%%%%%%%%%%%%%%%%%
%%%%%%%%%%%%%%%%%%%%%%%%%%%%%%%%%%%%%%